\documentclass[10pt,journal,compsoc]{IEEEtran}
\ifCLASSOPTIONcompsoc
  \usepackage[nocompress]{cite}
\else
  \usepackage{cite}
\fi
\ifCLASSINFOpdf
\usepackage[pdftex]{graphicx}
\graphicspath{{figures/}}
\else
\usepackage[dvips]{graphicx}
\graphicspath{{figures/}}
\fi
\usepackage{amsmath}
\usepackage{amsfonts}
\usepackage[ruled,norelsize]{algorithm2e}
\ifCLASSOPTIONcompsoc
 \usepackage[caption=false,font=footnotesize,labelfont=sf,textfont=sf]{subfig}
\else
 \usepackage[caption=false,font=footnotesize]{subfig}
\fi

\usepackage{wrapfig}

\usepackage[hidelinks]{hyperref}
\hypersetup{
    colorlinks=false,
    linkcolor=blue,
    filecolor=magenta,      
    urlcolor=cyan,
    pdftitle={CreatureShop},
}
\usepackage{caption}

\begin{document}

%
\title{CreatureShop: Interactive 3D Character Modeling and Texturing\\from a Single Color Drawing}
\IEEEaftertitletext{\vspace{15em}}

%
%
%
%

\author{Congyi~Zhang,
        Lei~Yang,
        Nenglun~Chen,
        Nicholas~Vining,\\
        Alla~Sheffer,
        Francis~C.M.~Lau,
        Guoping~Wang,
        Wenping~Wang
\IEEEcompsocitemizethanks{
\IEEEcompsocthanksitem
Congyi Zhang, Lei Yang, Nenglun Chen and Francis~C.M.~Lau are with the University of Hong Kong. Email: cyzhang@cs.hku.hk, lyang@cs.hku.hk, nolenc@hku.hk, fcmlau@cs.hku.hk.
\IEEEcompsocthanksitem
Nicholas Vining is with both NVIDIA and the University of British Columbia. Email: nvining@cs.ubc.ca
\IEEEcompsocthanksitem
Alla Sheffer is with the University of British Columbia. Email: sheffa@cs.ubc.ca
\IEEEcompsocthanksitem
Guoping Wang is with Peking University. Email: wgp@pku.edu.cn.
\IEEEcompsocthanksitem
Wenping Wang is with Texas A\&M University. Email: wenping@tamu.edu.
}
\thanks{Manuscript received April, 2022; revised July, 2022.}
}

\IEEEtitleabstractindextext{%
\begin{abstract}

Creating 3D shapes from 2D drawings is an important problem with applications in content creation for computer animation and virtual reality.
We introduce a new sketch-based system, {\em CreatureShop}, that enables amateurs to create high-quality textured 3D character models from 2D drawings with ease and efficiency.
CreatureShop takes an input bitmap drawing of a character (such as an animal or other creature), depicted from an arbitrary descriptive pose and viewpoint, and creates a 3D shape with plausible geometric details and textures from a small number of user annotations on the 2D drawing.
Our key contributions are a novel oblique view modeling method, a set of systematic approaches for producing plausible textures on the invisible or occluded parts of the 3D character (as viewed from the direction of the input drawing), and a user-friendly interactive system.
We validate our system and methods by creating numerous 3D characters from various drawings, and compare our results with related works to show the advantages of our method. We perform a user study to evaluate the usability of our system, which demonstrates that our system is a practical and efficient approach to create fully-textured 3D character models for novice users. 
\end{abstract}

\begin{IEEEkeywords}
character modeling, character texturing, interactive techniques
\end{IEEEkeywords}}

\maketitle
\begin{minipage}[t]{0.94\textwidth}
    \vspace{-17.5em}
    \includegraphics[width=\textwidth]{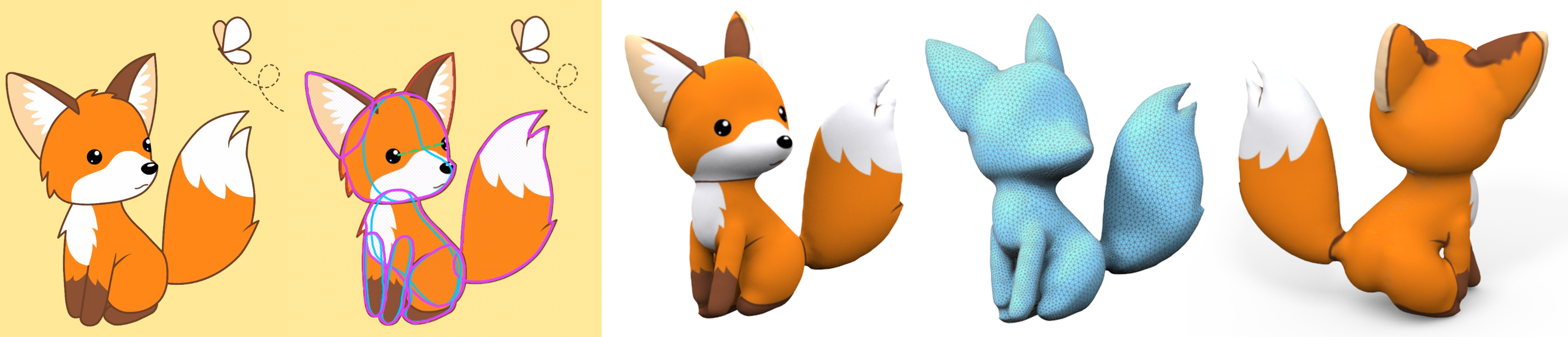}
    \captionof{figure}{CreatureShop converts arbitrary bitmap character drawings (left), using a set of intuitive user annotations (second left), into watertight textured 3D meshes (right). Fox by Natalia Linnik (\url{https://www.twitter.com/natalia_linn}); used with permission.}\label{fig:teaser}
\end{minipage}

\IEEEdisplaynontitleabstractindextext

\IEEEraisesectionheading{\section{Introduction}}

\IEEEPARstart{W}{ith} recent advances in virtual reality and bespoke manufacturing, it is increasingly desirable to empower amateur users to create compelling digital content. Character creation, in particular, has applications ranging from animation and games to toy manufacturing and virtual environments \cite{Borosan2012}\cite{Feng2017VR}\cite{Dvoroznak2020TOG}.

Creating a 3D character involves two major steps: modeling the 3D shape of the character, and defining its surface texture. Unlike professional modeling tools such as 3DS MAX \cite{3DSMAX} or ZBrush \cite{ZBrush}, content creation tools for novice users are usually domain-specific and aim to make their operation as intuitive as possible. Diverse modeling tools have been proposed for modeling character shapes, either using free-hand sketching interfaces \cite{Igarashi1999, Nealen2007, Borosan2012} or by taking advantage of existing 2D drawings \cite{Gingold2009, Bessmeltsev2015, ENTEM2015, Dvoroznak2020TOG}. Many separate solutions have also been proposed for texturing 3D models; several interactive tools \cite{Igarashi2001, Zhou2005, Schmidt2006} focus on painting the surface texture directly \cite{Igarashi2001}, or synthesize a surface texture from multiple images capturing different views of the object \cite{Zhou2005}. While these methods produce compelling texture on 3D shapes in the hands of professional artists, they rely on the artistic skill of the users or assume complete surface coverage by multiple-view images. 

Exploiting 2D drawings for 3D character creation offers numerous advantages. Perhaps the most notable advantage of 2D drawings is that they can specify both a character's shape and its surface texture at the same time. Characters are also typically drawn with poses and in views that convey a maximal amount of information about the character's shape and appearance, allowing viewers to ideate the drawn character from novel poses and views. 2D drawings also allow users to create characters from a wealth of legacy content, and to create characters by first drawing or painting with traditional media and then digitizing the results using a consumer scanner or phone. 

Many existing techniques for converting character drawings to 3D shapes \cite{ENTEM2015, Feng2017VR, Ramos2018CG, Dvoroznak2020TOG} only operate on characters drawn orthogonally from the side; other approaches \cite{Gingold2009} let users create characters based on 2D drawings with arbitrary views by defining a series of simple geometric proxies, then letting users manipulate them to approximately represent reference drawings. The resulting shapes do not produce high-fidelity reconstructions that represent the input drawings faithfully, and are therefore not suitable for downstream applications such as 3D printing or digital media. Generally speaking, previous methods either generate character surface geometry without considering surface texture, or obviate the need for sophisticated texture reconstruction by exploiting the requirement that characters are drawn from the side view only. In both cases, these methods fail to take full advantage of the descriptive ability of 2D character drawings. Creating a fully textured character from a drawing from an arbitrary descriptive view remains an open problem.

To address this problem, we present {\em CreatureShop}, an interactive system that enables amateur users to construct varied creatures and characters from a single input drawing of a reference character drawn from a descriptive viewpoint. As adjusting a 3D viewport is considered demanding for amateur users \cite{Gingold2009}, we restrict user interactions to a single view and image, and only require that the user annotates the drawing with a set of simple 2D annotations (Fig. \ref{fig:pipeline_a}); from these annotations, we generate complete, plausible surface shapes and textures that are consistent with the reference image.

A key challenge in modeling individual creature body parts, or {\em components}, from a single drawing is that only the outline of the component is provided; these outlines are inherently ambiguous and often non-planar \cite{Bessmeltsev2015}. We observe that bilateral symmetry in creature images, specifically visual clues such as eyes and noses, serves as a crucial cue that viewers use when ideating both the shape and orientation of individual creature body parts. We therefore design a set of simple user tasks to identify these cues in an input drawing, and then solve for bilaterally symmetric shapes with projected drawing-plane constraints given by part outlines; we can then exploit these outlines and annotations further in order to produce plausible surface textures for the geometry of individual components (Fig. \ref{fig:pipeline_b}). After modeling and texturing individual components in a part-based fashion, we then merge the textured components into a watertight character model, and combine multiple texture patches from individual components to produce a seamless texture over the character surface geometry with smooth transitions. As some portions of the creature are invisible in the input drawing, they may not have texture information that can be directly leveraged; we therefore use a surface-based inpainting technique to tackle this challenge, which generates seamless textures on occluded or invisible texture regions using context from faithful regions (Fig. \ref{fig:pipeline_c}).

Our technical contributions are as follows:

\begin{enumerate}
    \item We propose a character modeling method that can create creature geometries with details from a single reference image using a minimal set of 2D annotations;
    \item We show how to automatically produce plausible seamless textures for the generated creature models by leveraging the 2D texture of the input drawing;
    \item Finally, we design a user-friendly interactive system that integrates the modeling and texturing steps to create a fully textured character from a single reference drawing.
\end{enumerate}

As shown in our results, CreatureShop fulfills our design goals of taking bitmap drawings, leveraging simple user tasks, and producing high-quality character geometry with visually pleasing textures. We evaluate the performance of CreatureShop by showing a number of models created from color drawings by artists of various skill levels and using various media. We validate our approach by a pilot user study, which shows that our chosen interactions are natural and intuitive and that our generated models agree with viewer expectations. We also exhibit high-quality, watertight models made by novice users with no prior modeling experience within 14 minutes using our system. We compare our work to prior automatic and semi-automatic systems for character modeling from images, producing models of comparable or better quality from their input drawings and showing how their systems fail on inputs that CreatureShop successfully handles. Finally, since we generate textured watertight meshes, we show colored 3D printed toys created from drawings by our system.

\section{Related Work}
Our work is related to three areas of previous work: single-view character modeling, character texturing, texture inpainting and generation.

\emph{Single-View Character Modeling.} The key challenge when generating a character model from a single character drawing is that arbitrary character drawings can be extremely ambiguous or open to multiple interpretations; producing a valid character model requires resolving these ambiguities.

Prior knowledge is often useful for specific tasks, or for assisting sketch-based modeling. \cite{Kanazawa2016} models 3D animals from user-clicked single images by fitting 3D mesh templates to drawings. The modeling category is confined to the animal template being used, which makes it hard to extend to more general cases in character drawings. \cite{Li2018} uses a convolutional neural network model to process both single- and multi-view sketches. For the single view case, their method can create faithful surfaces patches for the original view, but cannot complete the back side to compose a watertight model. \cite{Kanazawa2018} recovers a single 3D object from its single-view image and produces coarse shape and texture on it. A recent neural rendering based symmetric geometry inference method has been proposed to generate depth maps for realistic single-view photos \cite{Wu2020CVPR}. Complete and complex shapes and delicate textures are still hard to generated with learning-based methods, especially for non-realistic drawings.

Non-learning based approaches use domain-specific assumptions to resolve such ambiguity, as shown in \cite{Bessmeltsev2015}. Assumptions are typically made to restrict the shapes under consideration - e.g. mirror-symmetric shapes in \cite{Cordier2011}, sweeping surfaces in \cite{Andre2011}, and generalized cylinders as geometric primitives in \cite{Gingold2009} and \cite{Bessmeltsev2015}. Further assumptions are made in different approaches, such as limiting the input drawings to frontal views \cite{Buchanan2013} or side views \cite{ENTEM2015,Dvoroznak2020TOG}, or assuming that the silhouette of the body part is planar and perpendicular to the view direction \cite{Karpenko2006,Cordier2011}. Our system requires no specific views in the input drawings and makes no assumption of planar silhouettes. Instead, we exploit mirror symmetry to simplify the user interactions required to create organic parts with fine details.

Our method is similar to prior art (e.g. \cite{Gingold2009}, \cite{Olsen2011}, \cite{Andre2011}, \cite{Yeh2017}) in that it asks users to outline parts of the drawing to denote component elements; we further ask users to provide additional strokes with respect to the reference drawings, since these additional inputs provide useful information for both modeling and texturing. The key difference between the modeling section of our method and Gingold et al. \cite{Gingold2009} is that their method solely creates generalized cylinders and ellipsoids; our created parts are more general and directly agree with the original outline specified by the user, improving geometric fidelity. Olsen et al. \cite{Olsen2011} assume that user strokes and consequently part outlines always lie on the fixed canvas drawing plane; by allowing users to specify part midlines and symmetry points, we do not have any such restriction. The tool developed by Andre and Saito \cite{Andre2011} requires users to draw certain cross-sectional profiles of the shape satisfying specific geometry relationships; manually specifying these geometric relationships is demanding for novice users. Similar curves are used in CreatureShop, but the geometric relationship is automatically determined by our modeling system rather than by users. Yeh et al. \cite{Yeh2017} provide a set of interactive tools allowing users to create 2.5D high relief models from photos. Unlike CreatureShop, the outputs of their method are not fully textured 3D models, and are thus not suitable for 3D printing or other downstream applications.

\emph{Character Texturing.} Many previous works on 3D model creation from reference drawings have presented their models with textures; however, the texture is directly projected onto the surface (\cite{Olsen2011,Buchanan2013, Bessmeltsev2015,Dvoroznak2020TOG}) resulting in distorted textures and poor quality results. \cite{Buchanan2013} expands the boundary of the texture region for frontal view drawings to resolve the distortion caused by this parallel projection approach, but the success of this method largely depends on texture complexity near the boundary. Recently, a category-specific learning-based automatic modeling and texturing method has been proposed \cite{Zuffi2019ICCV} which estimates the pose, shape, and plausible texture of zebras in photographs; it is unclear how this method can be generalized to arbitrary drawings. We propose a general approach for texture mapping that alleviates distortion near outlines for arbitrary views.

Another category of character texturing work lets users paint textures onto 3D surfaces interactively \cite{Igarashi2001}\cite{Carr2004}, to create texture maps for models using given decals \cite{Schmidt2006}, or use multi-view images for texture synthesis \cite{Zhou2005} \cite{Gal2010}. Those approaches require additional workflow following model creation. To the best of our knowledge, no previous work has addressed the problem of simultaneously extracting a character model and texture from a single input drawing. Additionally, no previous work that we are aware of has explored how best to generate plausible textures for occluded and back-facing regions of models from character drawings when only a single view is available. To the best of our knowledge, an integrated pipeline for seam repair as well as texture completion using the same user input as the modeling phase has not been devised yet.

\emph{Image inpainting and texture synthesis.} Our surface texturing method is related to the image inpainting problem in patch based synthesis; see \cite{Barnes2017} for a recent survey. \cite{Darabi2012} is capable of merging two source images or complete missing regions in a image using a content-aware manner based on PatchMatch\cite{Barnes2009}. Direct use of image inpainting methods requires a planar parameterization that contains the target region, which would produce either distortion or discontinuity, especially in regions with complex geometry. This motivates our approach, which performs inpainting on the manifold surface rather than the image domain. \cite{Turk2001}\cite{Wei2001} both extend texture synthesis algorithms from rectangular domains to arbitrary manifold surfaces, and \cite{Chen2012} transfers properties such as color texture between two surfaces using a mesh-based PatchMatch algorithm; however these methods all discretize texture information to colors on mesh vertices, resulting in low resolution inpainting unless the number of vertices on the mesh is sufficiently high. Consequently, the newly generated texture will be blurry and will have obvious seams between the visible and inpainted regions. To address this, we develop a surface-based inpainting to generate seamless high resolution results without increasing time consumption.
\section{System Overview}

\begin{figure*}[!t]
\centering
\subfloat[User annotations]{\includegraphics[height=0.22\linewidth]{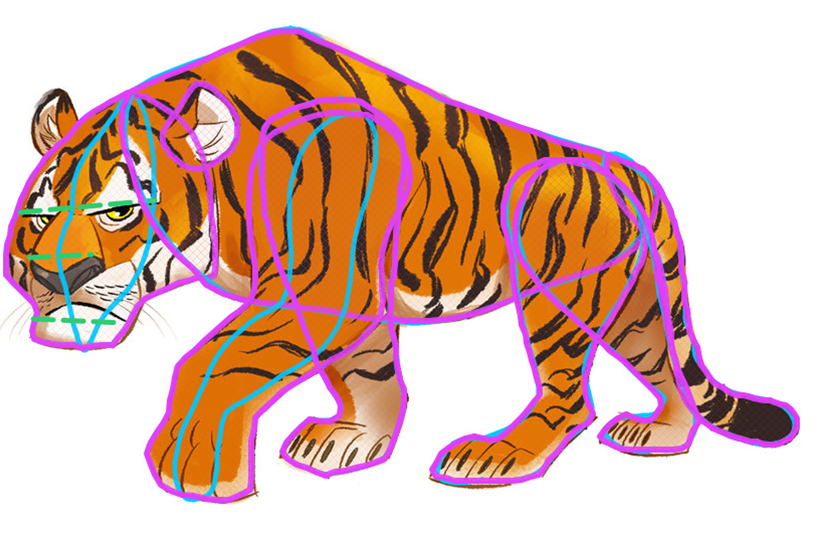}%
\label{fig:pipeline_a}}
\hfil
\subfloat[Components and their symmetry planes]{\includegraphics[height=0.22\linewidth]{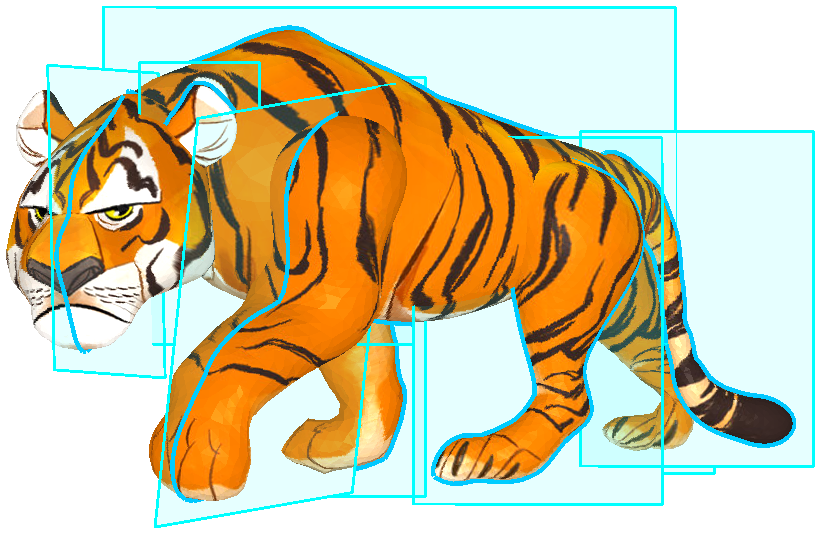}%
\label{fig:pipeline_b}}
\hfil
\subfloat[A novel view]{\includegraphics[height=0.3\linewidth]{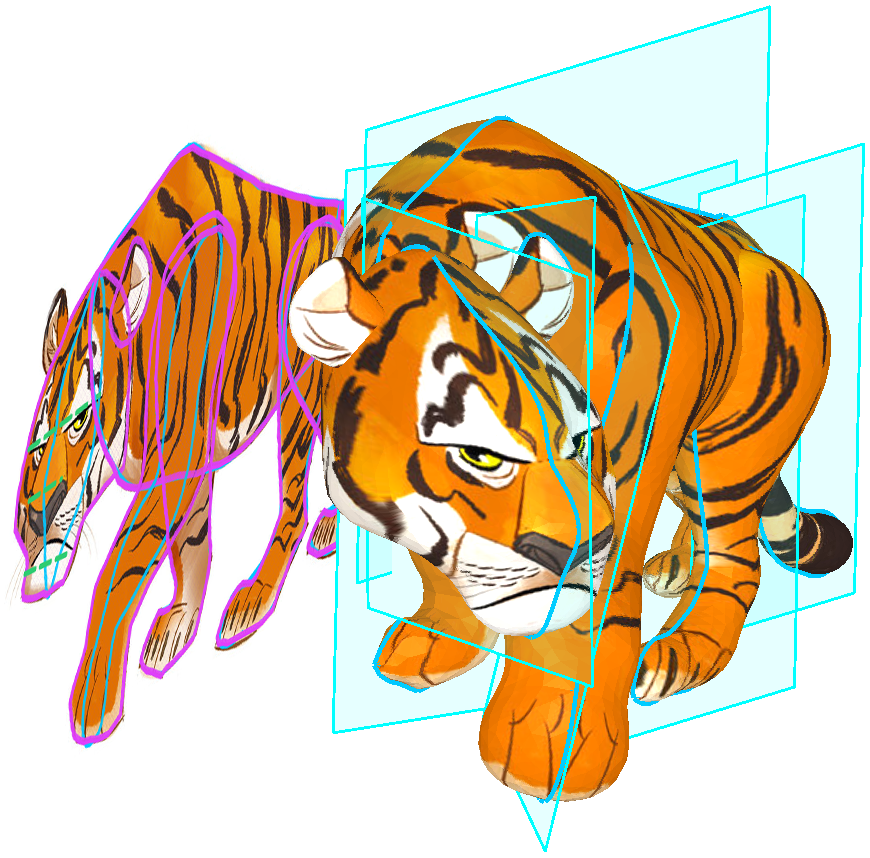}%
\label{fig:pipeline_c}}
\caption{CreatureShop workflow. Starting with a set of user annotations (a), CreatureShop produces a set of individual components (b) that are combined to create a fully textured character model (c, here seen from a novel view). Tiger obtained from \url{https://www.pngwing.com/en/free-png-zzzye}, under license for non-commercial use.}
\label{fig:pipeline}
\end{figure*}

CreatureShop takes a bitmap drawing and a series of user annotations as input, and produces a watertight character model with a plausible texture in a part-based manner. As lifting a 2D character from a single drawing to 3D is an ill-posed problem, the system relies on two assumptions to produce satisfactory results. First, we require that the input drawing is sufficiently descriptive; more specifically, we require that the input drawing presents a character from an oblique or side view with an orthogonal projection, and that the target character is not drawn in the frontal view. Second, we assume that the output character has bilaterally symmetric geometry and texture, in order to allow us to make the best use of the textural information contained in the input drawing. In what follows, we first introduce the four user annotations (and tasks) that are provided in the interface for creating individual components with textures, and then present a general pipeline to produce a textured character model.

\subsection{User Annotations.}

CreatureShop provides four simple, intuitive user annotation tools that are applied to the 2D drawing plane to create individual textured components; we then leverage these annotations and components to create the final 3D textured character. We allow the user to perform four basic annotations: component outlining; specifying the orientation of the symmetry plane; identifying symmetric component features; and defining the midline. The first two operations create basic shapes, and the second two are used to refine the basic shape.

\emph{The outline of a component}, denoted $o$, is the planar curve that bounds the region occupied by the component in the drawing (purple lines in Fig.\ref{fig:pipeline_a}). Component outlines are used to confine the projection of the overall shape of the component to be in correspondence with the reference shape in the input drawing. The outline of each component is depicted by refining the outline of the overall character in our user interface.

\emph{The orientation of the symmetry plane} defines the orientation of the component with which it is bilaterally symmetric. The orientation of the symmetry plane is useful for assigning symmetric textures by reflection, maximizing the use of the visible texture from the reference drawing, and for defining the geometry of each component.
Users can drag a line segment to imply the normal direction of the symmetry plane in the drawing plane.

\emph{Pairs of symmetric landmarks}, such as eyes, form important cues from the drawing for establishing bilateral symmetry. Users are quickly able to identify such landmarks in CreatureShop by linking a line segment to connect them (green dashed line in Fig.\ref{fig:pipeline_a}); we then exploit these pairs to create more details to the geometry.

\emph{The midline of a component}, denoted $m$, is the 2D projection, in the drawing plane (blue lines in Fig.\ref{fig:pipeline_a}), of the symmetry curve $M$ (blue lines in Fig.\ref{fig:pipeline_b}). The symmetry curve is intersected by the shape and its unique symmetry plane. The midline is useful for controlling local shape features around the symmetry plane of components, such as a protruding nose on a face, which are unique features in the symmetry plane and cannot be defined solely by symmetric landmarks. Users are able to adjust the initial shapes of midlines generated by our system and see real-time feedback in 3D view, which is presented next to the drawing canvas in our user interface.

\subsection{Pipeline}

Our algorithm can be divided into two stages, a modeling stage and a texturing stage. Both stages operate from the same input bitmap and user annotations; the modeling stage defines a set of body parts based on the input bitmap and user annotations, while providing real-time visual feedback as annotations are defined. The texturing stage, in turn, employs the same set of annotations as the modeling stage to define texture information for the entire creature, including areas of the creature where texture information is occluded or otherwise not present in the original drawing.

The major steps taken to create a fully textured character in CreatureShop are as follows:

\textbf{User Part Annotation.} Given an input drawing satisfying our descriptive view assumption, the user starts by extracting the overall character region from the input drawing, and then segments each body part of the character. The result of this stage is a set of partially overlapping subregions.

\textbf{Creation of Individual Body Parts.} Given a sparse set of planar annotations with different semantics, we formulate the creation of plausible 3D geometry as a constrained optimization problem. Using the outline as a constraint, the user can orient the symmetry plane for each body part to produce components in a satisfactory pose and with plausible textures. Pairs of symmetric landmarks and midlines can be identified and adjusted interactively at this stage, as the user wishes, to add geometric details to the components (Section 4.2). Each individual body part is then automatically textured using the previously specified user annotations (Section 4.3).

\textbf{Character Composition with Complete Texture.} Finally, we assemble components and assign texture information to the complete creature to produce the final model. 
For individual body parts, we provide users two ways to specify their desired depths and symmetry planes: individually or as a pair (for parts with intrinsic symmetry).
For body parts that are created in pairs (e.g. legs), we also transfer the texture from the body part at the near-view side to that at the far-view side, so as to ensure the indicated part symmetry during modeling (Section 5.2). Users may further adjust these default estimations as they wish.

The well-positioned body parts are then unified into a seamless, watertight mesh. However, there are still some ill-textured regions to be handled due to self-occlusion or occlusion between components. The system automatically detects and fills these ill-textured regions (Section 5.1). The resulting output is a 3D character model with a fully defined, plausible texture.
\section{Creation of a Single Body Part}\label{sec_single}

In this section, we focus on how a single component is created from the outline and additional annotations furnished by users.
We formulate this as a shape optimization problem that must reliably lift the user's 2D annotations to form a smooth surface mesh in 3D. We first describe the shape optimization framework, and then show how different annotations are incorporated to help shape the geometry. Finally, we introduce a simple yet versatile texturing scheme to produce reliable texture information on the visible side of the 3D surface mesh; this reliable texture information will then be extended over the entire creature during the final component merge phase.

\subsection{Shape Optimization Formulation}

CreatureShop uses a set of visual cues, such as body part outlines, to find a smooth shape to approximate the underlying component. As identified in the drawing, these visual cues are two-dimensional and have only $X$ and $Y$ coordinates; our first task is therefore to find the depths (or $Z$ values) for these visual cues. For simplicity and without loss of generality, throughout the paper we will define the drawing plane to be located at $Z=0$, and the $Z$-axis is assumed to be opposite to the view direction.

We derive a smooth, plausible and bilaterally symmmetric surface geometry by extending the Fibermesh shape optimization framework proposed in~\cite{Nealen2007}, which finds a smooth shape from a set of definite spatial curves constraints by interleaving smoothing surface curvatures. In our case, we extend this framework by adding a set of depth-variable spatial constraints with inherent symmetric shape constraints.
We first compose an initial discrete triangular mesh structure and then deform it to the final shape using our proposed optimization. The initial mesh structure is symmetric about the outline curve $o$; we generate one side of the mesh structure via constrained Delaunay triangulation using the outline $o$ as a boundary constraint. Specifically, we triangulate the target region with dense triangles ($>1600$), controlling those triangles to be uniform and as close to equilateral as possible by imposing a maximum triangle area constraint and a minimum triangle angle constraint. This triangulation strategy helps convergence when solving Fibermesh, which was observed to be unstable in certain conditions \cite{Andrews2011}. We then reflect the triangles across the symmetry plane $\pi$ (initialized as $Z=0$, the drawing plane) and join the boundaries together. We then inflate this shape using the Fibermesh algorithm. The collection of vertices lying in $\pi$ form a set $\mathcal{M}$ delineating the \emph{midline} of the shape. For all other vertices on one side of $\pi$ we can find corresponding vertices on the opposite side, forming symmetric pairs about $\pi$. The set of all pairs of symmetric vertices about $\pi$ is denoted $\mathcal{S}$. For efficiency, during the optimization, the mesh connectivity remains unchanged.

We use the first step of the optimization framework proposed in \cite{Nealen2007} to solve Laplacian magnitudes $\{c_i\}$ and edge lengths $\{e_i\}$ both of which smoothly vary over the surface:

\begin{equation}\label{eq:curvature_edgelength}
\begin{split}
  &\arg\min_{c}\left\{\sum_{i}{\|\mathbf{L}(c_i)\|}^2+\sum_{i}{\|c_i-c^\prime_i\|}^2\right\}\\
  &\arg\min_{e}\left\{\sum_{i}{\|\mathbf{L}(e_i)\|}^2+\sum_{i}{\|e_i-e^\prime_i\|}^2\right\}
\end{split}
\end{equation}
where $\mathbf{L}(\cdot)$ denotes the discrete graph Laplacian, and $\{c_i^\prime\}$ and $\{e_i^\prime\}$ are the current Laplacian magnitudes and average edge length incident on vertex $i$ on surface respectively.

\begin{figure}[htbp]
    \centering
    \includegraphics[width=0.8\columnwidth]{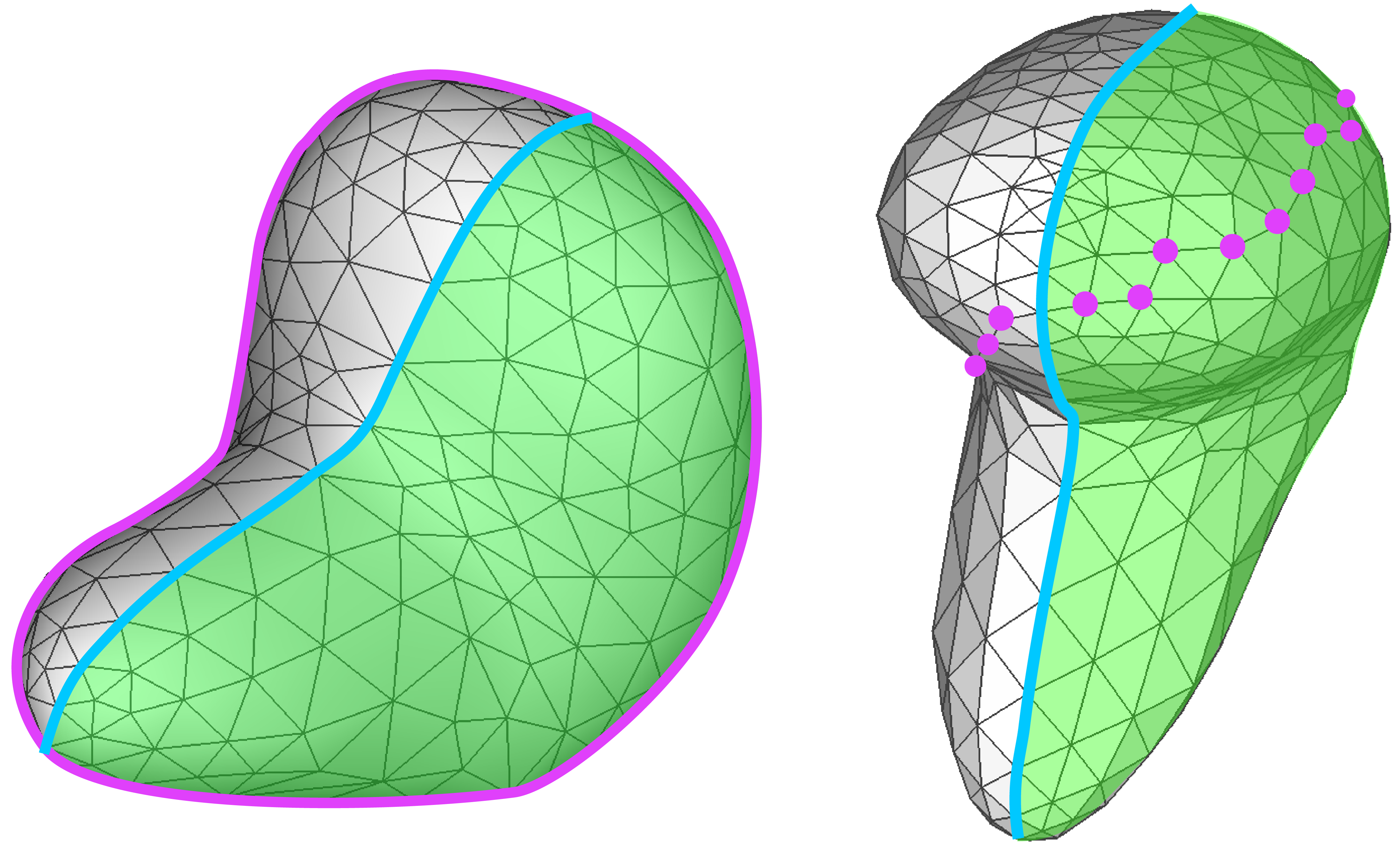}
    \caption{The mesh vertices in the green region forms the set $S$, each of which has a symmetric correspondence at the other side of the symmetry curve $M$ (blue line), which is composed by a closed loop of vertices lying on the symmetry plane $\pi$. The projection constraints set $B$, for example, are illustrated as purple vertices in the right figure; their projection on the drawing plane lies on the outline (purple line in the left).}
    \label{fig:mesh_connectivity}
\end{figure}

To account for the variability of the depth of the shape silhouette whose projection on the drawing plane is the outline, we augment this basic formulation with a set of symmetry-related constraints to recover the surface geometry along with the Laplacians and edge lengths solved above, as well as constraints derived from the user-specified planar annotations. The formulation of the second step yields:

\begin{equation}\label{eq:position_solver}
  \arg\min_{\mathbf{v}}\left\{\alpha_c E_c+\alpha_e E_e+\alpha_s E_s+\alpha_m E_m+\alpha_p E_p\right\}.
\end{equation}

In Eq.\ref{eq:position_solver}, the discrete Laplacians term $E_c$ and edge lengths term $E_e$ are chosen to ensure the smoothness and mesh quality of the surface mesh, respectively, as per \cite{Nealen2007}; they are defined as follows:

\begin{align}
    E_c & =\sum_{\mathbf{v}_i \in \mathcal{V}}{\|\mathbf{L}(\mathbf{v}_i)-\delta_i\|^2},\label{eq:e_c}\\
    E_e & =\sum_{(\mathbf{v}_i,\mathbf{v}_j)\in \mathcal{E}}{\|\mathbf{v}_i-\mathbf{v}_j-\eta_{ij}\|^2},\label{eq:e_e}
\end{align}
where $\delta_i=A_i\cdot c_i\cdot \mathbf{n}_i$, $A_i$ and $\mathbf{n}_i$ are area and normal estimates for the vertex $i$;
$\mathcal{V}$ is the set of all vertices, $\mathcal{E}$ the set of pairs of vertices connected by edges, and $\eta_{ij}=\frac{e_i+e_j}{2}\cdot\frac{\mathbf{v}^\prime_i-\mathbf{v}^\prime_j}{\|\mathbf{v}^\prime_i-\mathbf{v}^\prime_j\|}$.

The symmetry-related constraints include two terms, $E_s$ and $E_m$. The first term is a symmetric vertex pair constraint that ensures pairs of symmetric vertices residing on either side of the mesh remain symmetric during the modeling process. The second term is designed to encourage the midline vertices to stay in the symmetry plane during optimization.

\begin{align}
    E_s &= \sum_{(\mathbf{v}_i,\mathbf{v}_j)\in \mathcal{S}}{\|\mathbf{n}_\pi\cdot\frac{\mathbf{v}_i+\mathbf{v}_j}{2}+d\|^2 +
        \|\mathbf{n}_\pi\times(\mathbf{v}_i-\mathbf{v}_j)\|^2}, \\
    E_m &= \sum_{\mathbf{v}_i\in \mathcal{M}}{\|\mathbf{n}_\pi\cdot\mathbf{v}_i+d\|^2}.
\end{align}
where $\pi$ is the symmetry plane with equation $ax+by+cz+d=0$, $\mathbf{n}_\pi(a,b,c)$ is the unit normal of $\pi$,  $\mathcal{S}$ is the set of all symmetric pairs of vertices about $\pi$ in the mesh, and $\mathcal{M}$ is the set of vertices in the symmetry plane.

The last term in Eq.\ref{eq:position_solver} is a projection constraint term that incorporates the user-specified annotations in the drawing plane:
\begin{equation}\label{eq:e_p}
    E_p=\sum_{\mathbf{v}_i\in \mathcal{B}}{\|P(\mathbf{v}_i)-\bar{\mathbf{p}}_i\|^2}.
\end{equation}
where $\mathcal{B}$ is the set of projection constraints, $\bar{\mathbf{p}}_i$ are the target positions in the drawing plane of the vertices and $P$ is the orthogonal projection that maps $\mathbf{v}(v_x, v_y, v_z)\in \mathbb{R}^3$ to the drawing plane, i.e. $P(\mathbf{v}) = (v_x, v_y)$.

The linear least squares problems Eq.\ref{eq:curvature_edgelength} and Eq.\ref{eq:position_solver} can be optimized by solving two linear systems separately. We alternately minimize those two objective energies to obtain a result shape.

\begin{figure}[htbp]
\centering
\subfloat[]{\includegraphics[width=0.3\linewidth]{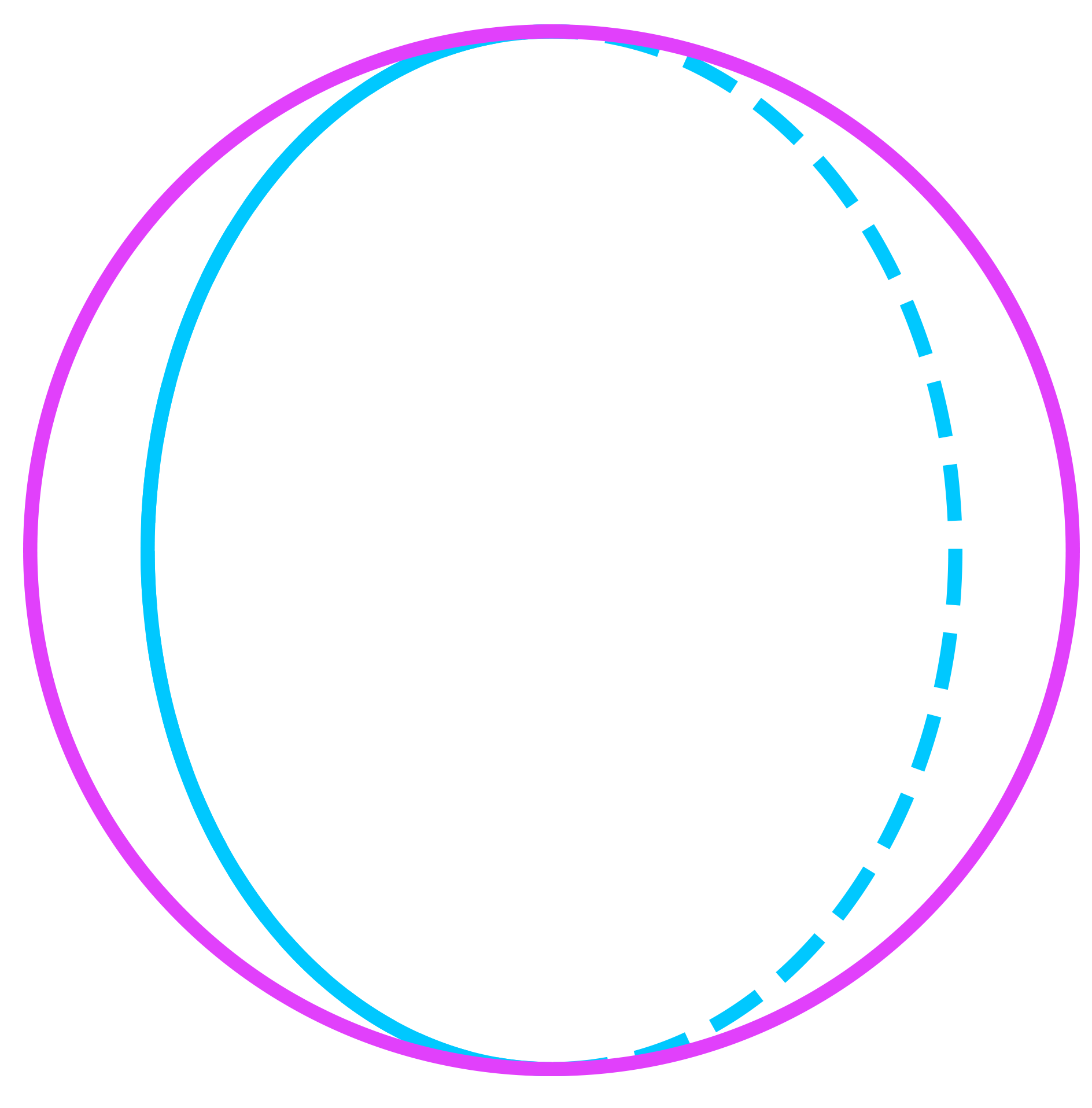}%
\label{fig:roundness_a}}
\hfil
\subfloat[$k=1$]{\includegraphics[width=0.3\linewidth]{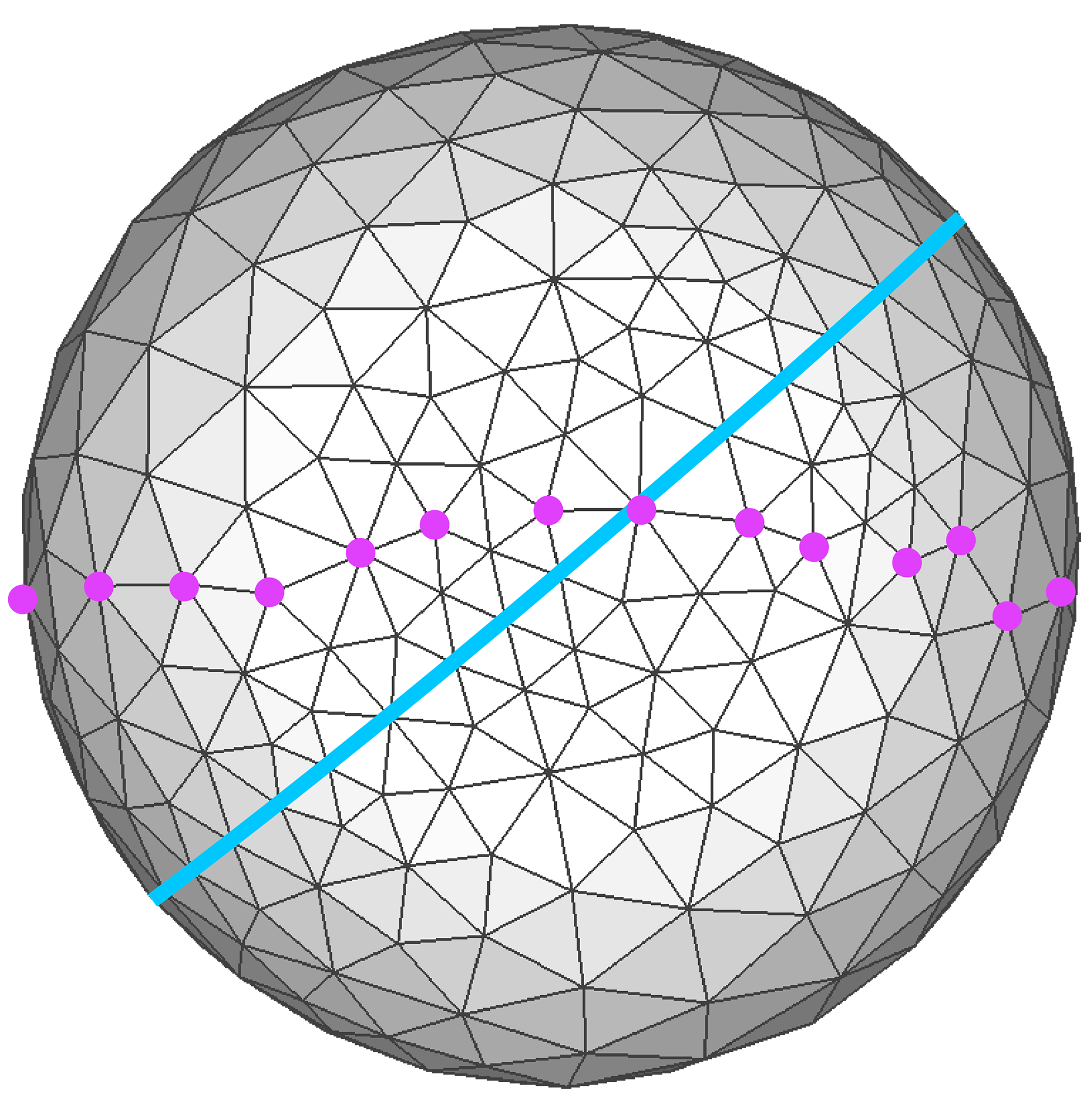}%
\label{fig:roundness_b}}
\hfil
\subfloat[$k=0.1$]{\includegraphics[width=0.3\linewidth]{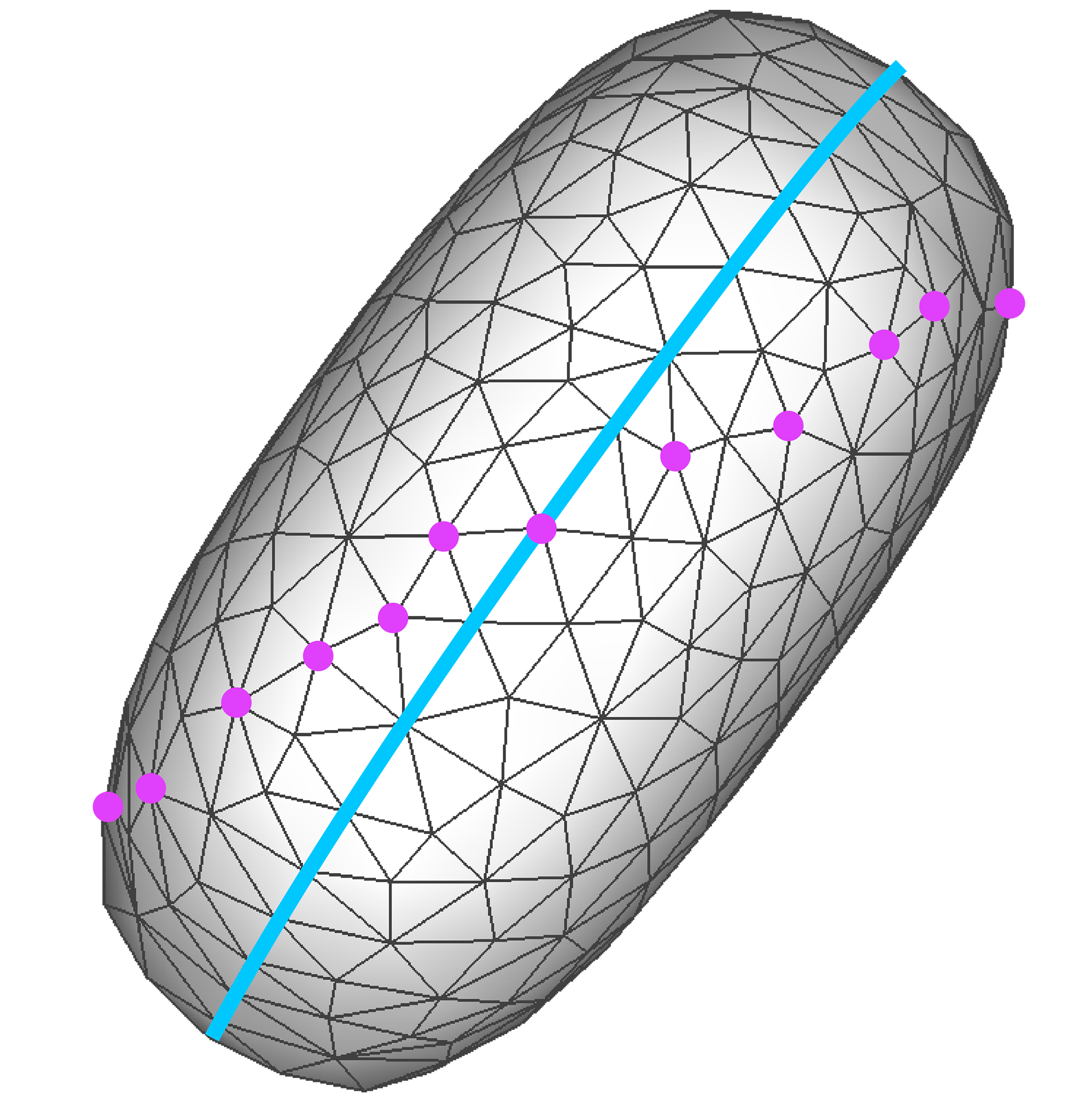}%
\label{fig:roundness_c}}
\caption{The effect of the thickness term $k$. (a) shows the outline and midline on drawing plane (front view). (b) and (c) are top views of the result shape using different values of $k$, symmetry curves are in blue lines, projection constraints are in purple points.}
\label{fig:roundness}
\end{figure}

All $\alpha$'s in Eq.\ref{eq:position_solver} are coefficients to balance those energies. We use fixed values of $\alpha_e=1$ and $\alpha_s=\alpha_m=\alpha_p=1000$ for all our examples; $\alpha_c=k\frac{\#\mathcal{E}}{\#B}$ is estimated dynamically relative to the mesh connection and projection constraints.
We allow users to adjust $k$ to control the thickness of results as shown in Fig.\ref{fig:roundness}.

As the mesh connectivity is unchanged after deformation, the pre-factorized matrices in the least-squared system of Eq.\ref{eq:curvature_edgelength} can be re-used with new constraints in an interactive rate. We found that alternating between Eq.\ref{eq:curvature_edgelength} and Eq.\ref{eq:position_solver} for approximately 5 iterations quickly converges to a satisfactory solution.

\subsection{Modeling the Surface Geometry}

\subsubsection{Creating base shapes with the outline and symmetry plane}

Based on our proposed method, modeling a base shape requires only the outline and a well-posed symmetry plane $\pi$ given the user. To help users specify the symmetry plane easily, CreatureShop adopts the following user interaction strategy. The system initially creates a symmetric surface mesh with the outline and the symmetry plane at $z=0$, in which case the optimization formulation is equivalent to FiberMesh \cite{Nealen2007}. The user is then asked to rotate the symmetry plane (and thus the shape) to a new pose from the drawing plane $z=0$ to approximate the depicted body part in the 2D drawing.

Given the properly rotated shape and its symmetry plane, we seek an optimal sequence of one-to-one matchings between points from the outline $\{\bar{\mathbf{p}}_i\}$
and the mesh vertices $\{\mathbf{v}_j\}$ from the shape silhouette. We find this sequence by employing a Hidden Markov Model approach, as proposed in \cite{Kraevoy2009}. We then enforce the requirement that vertices in the optimal sequence $\{\mathbf{v}_i^o\}$ must have their projections in the drawing plane to be $\{\bar{\mathbf{p}}_i\}$ by adding projection constraints using Eqn. \ref{eq:e_p} during the meshing process, with $\mathbf{v}_i$ as 3D vertices and $\bar{\mathbf{p}}_i$ as their 2D targets.

\subsubsection{Refining base shapes via user annotations}

The base shape created by the previous method only makes use of the outline and the orientation of the symmetry plane, and thus lacks details to enhance the underlying body part. Other user-specified cues, such as the symmetric landmarks and the shape of the midline, are valuable when defining the user's intended body part shapes. To this end, we allow users to incorporate these cues as additional constraints for shape refinement.

\paragraph{Symmetric landmarks as constraint}

Given a specified symmetry plane $\pi$ and a pair of symmetric landmarks, $\bar{\mathbf{p}}_1$ and $\bar{\mathbf{p}}_2$, in the drawing plane, we can easily find the nearest pair of symmetric vertices $\mathbf{v}_i$ and $\mathbf{v}_j$ on the mesh that best approximates the 3D positions $\mathbf{p}_1$ and $\mathbf{p}_2$ of the landmark points. (We note that Bae \textit{et al.} \cite{Bae2008} provides a single-view symmetric curve epipolar sketching tool for expert designers by adopting similar geometric principles to recover 3D curves from single-view 2D curves.) In our method, 3D coordinates of symmetric landmarks are solved in a least-square manner that both enforces the symmetry and accounts for any imprecision made by users or in the reference drawing.

\begin{equation}\label{eq:sym_pairs}
\begin{split}
  \arg\min_{\mathbf{v}_i,\mathbf{v}_j}\{&\|\mathbf{n}_\pi\cdot\frac{\mathbf{v}_i+\mathbf{v}_j}{2}+d\|^2 + \|\mathbf{n}_\pi\times(\mathbf{v}_i-\mathbf{v}_j)\|^2\\
  &+\|P(\mathbf{v}_i)-\bar{\mathbf{p}}_1\|^2+\|P(\mathbf{v}_j)-\bar{\mathbf{p}}_2\|^2\}
\end{split}
\end{equation}
By adding the indices $i$ and $j$ of $\mathbf{v}_1$ and $\mathbf{v}_2$ to the projection constraint set $\mathcal{B}$, we can solve for the refined shape by Eq. \ref{eq:position_solver}.

\paragraph{Midline as constraint}

The midline $m$ provides important information for inferring the underlying shape, but it is well-established \cite{Schmidt2009} that even expert artists have a hard time drawing an accurate midline for a shape in a single stroke; we therefore allow users to reshape the underlying geometry by interactively editing the midline in the drawing plane. This amounts to specifying the projected coordinates of the vertices in the symmetry curve, which again contributes to the projection constraint set $\mathcal{B}$. We note that the in-the-symmetry-plane term $E_m$ only restricts vertices in $\mathcal{M}$ to reside in the symmetry plane; thus the additional midline constraint and term $E_m$ serve as complementary terms for yielding the user's desired 3D shapes.

\begin{figure}[htbp]
\centering
\subfloat[]{\includegraphics[width=0.24\linewidth]{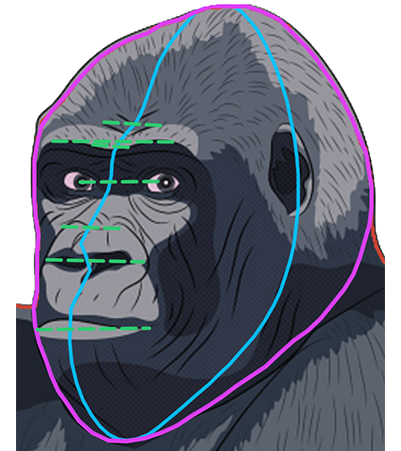}%
\label{fig:modeling_with_detail_a}}
\hfil
\subfloat[]{\includegraphics[width=0.24\linewidth]{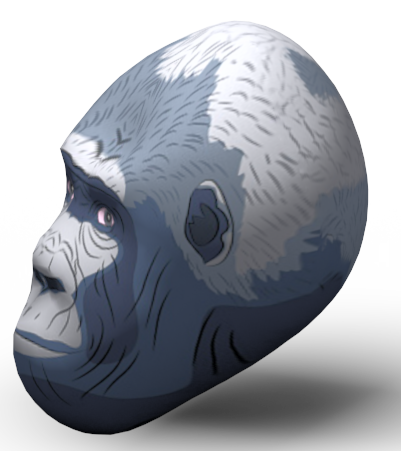}%
\label{fig:modeling_with_detail_b}}
\hfil
\subfloat[]{\includegraphics[width=0.24\linewidth]{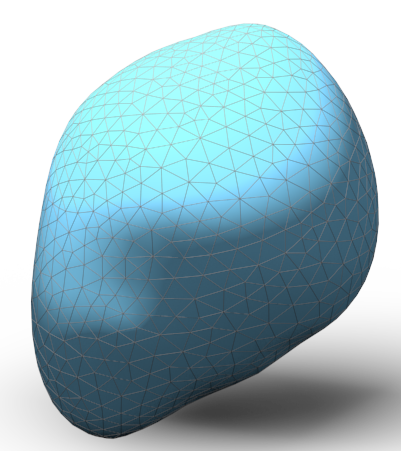}%
\label{fig:modeling_with_detail_c}}
\hfil
\subfloat[]{\includegraphics[width=0.24\linewidth]{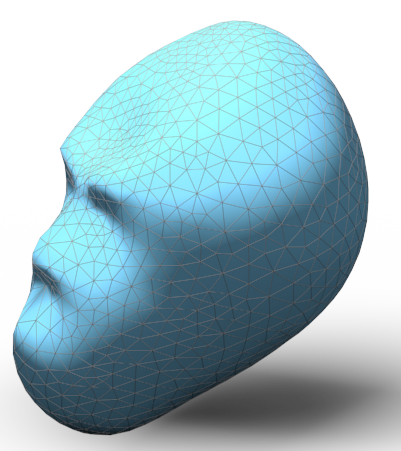}%
\label{fig:modeling_with_detail_d}}
\caption{Modeling a gorilla's head with an outline and additional inputs. The midline and the symmetric features largely enhance the visual quality of the geometry. In the reference drawing, the outline, midline and symmetric pairs are edited by the user (a). The final textured shape is shown in (b). The symmetry plane is rotated with only the outline constraint to produce the base shape (c). The base shape is then refined by editing the midline and adding several pairs of symmetric landmarks, as shown in (d). Gorilla from \url{https://www.vectorstock.com}, purchased under standard license.}
\label{fig:modeling_with_detail}
\end{figure}

As shown in Fig.\ref{fig:modeling_with_detail}, the midline can be used to reshape the geometry near the symmetry plane, while adding symmetric landmarks can enhance details of the geometry.

\subsubsection{Manipulating the orientation of the symmetry plane}

So far, we have assumed that a reasonable orientation of the symmetry plane is given for modeling a 3D shape by the outline and a few pairs of symmetric landmarks. However, accurately specifying the orientation of a 3D plane is challenging for novice users in the modeling system.
To obtain a reasonable orientation for the symmetry plane, we assume that the symmetry plane can be found by rotating the drawing plane. The applied rotation can be represented by an axis and an angle of rotation; the axis of rotation can be inferred from the drawing, and the user then only needs to specify the rotation angle by dragging the mouse. (Please see the supplementary video to see this system in action.)

\begin{figure}[htbp]
  \centering
  \includegraphics[width=0.43\columnwidth]{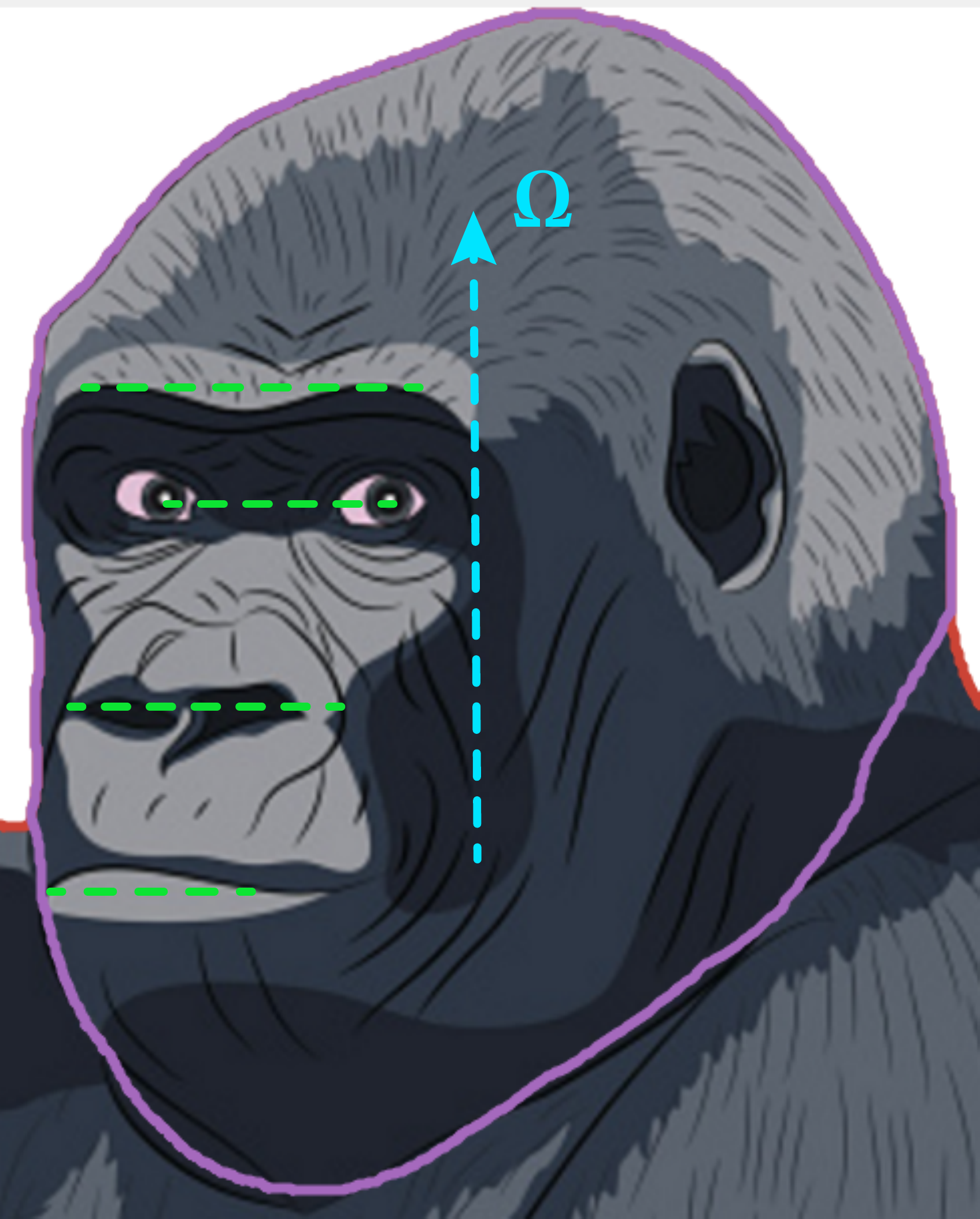}
  \caption{The rotation axis (shown in blue) is approximately orthogonal to the feature pairs (shown in green) in the drawing.}\label{fig:oblique_notations}
\end{figure}

Our system uses a simple geometric relationship to find the rotational axis $\Omega$. On the one hand, $\Omega$ must also be the intersection line that lies in both planes; thus its direction can be represented as $(\Omega_x,\Omega_y,0)$. On the other hand, as we assume that the symmetry plane can be found by rotating the drawing plane $z=0$ around $\Omega$, it follows that $\Omega$ must always be perpendicular to the symmetry plane normal $\mathbf{n}_\pi$, and thus to $P(\mathbf{n}_\pi)$, the projection of $\mathbf{n}_\pi$ in the drawing plane. $P(\mathbf{n}_\pi)$ can be determined easily from the direction of any pair of symmetric landmarks, such as a pair of eyes, in the reference drawing; the rotational axis $\Omega$ perpendicular to it and passing through the center of the outline is then uniquely defined (see Fig. \ref{fig:oblique_notations}).

Since determining the depth information with 2D planar constraints is an ill-posed task, the rotation angle $\theta$ must be inferred by the user by dragging the mouse.

CreatureShop provides users with an interactive way to estimate the rotation angle $\theta$. The estimation is achieved by asking users to drag a line whose direction is orthogonal to the rotational axis of the symmetry plane and whose length reflects the magnitude of the rotational angle. In this way, the user is able to interactively rotates the symmetry plane by dragging the line in the drawing plane, and is provided with a real-time visualization of the projection of a rotated circle (please refer to the demo of modeling process in the supplementary video).

\subsection{Texturing the individual component}

\begin{figure}[htbp]
\centering
\subfloat[]{\includegraphics[height=0.42\linewidth]{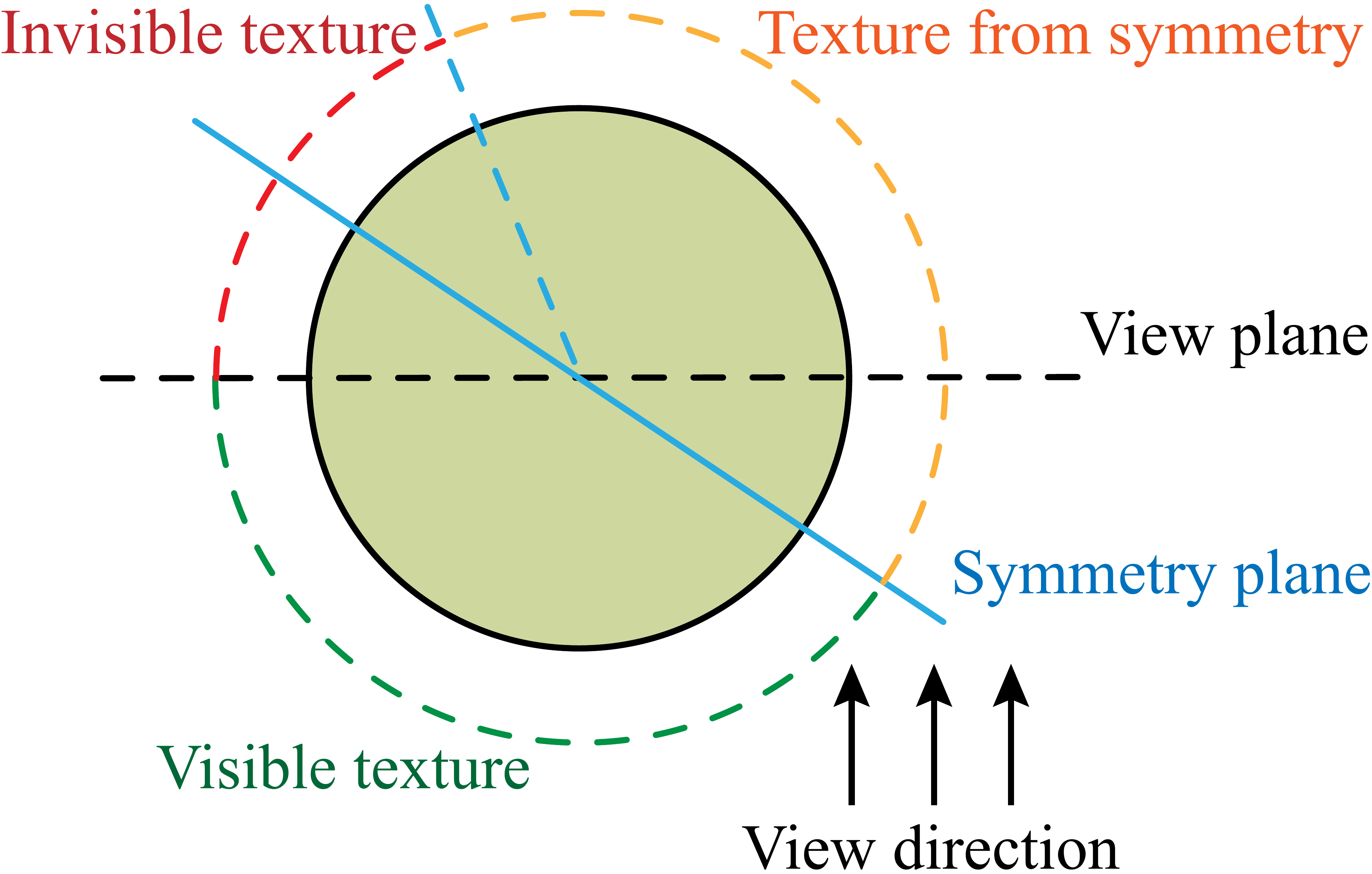}%
\label{fig:texture_visibility_a}}
\hfil
\subfloat[]{\includegraphics[height=0.42\linewidth]{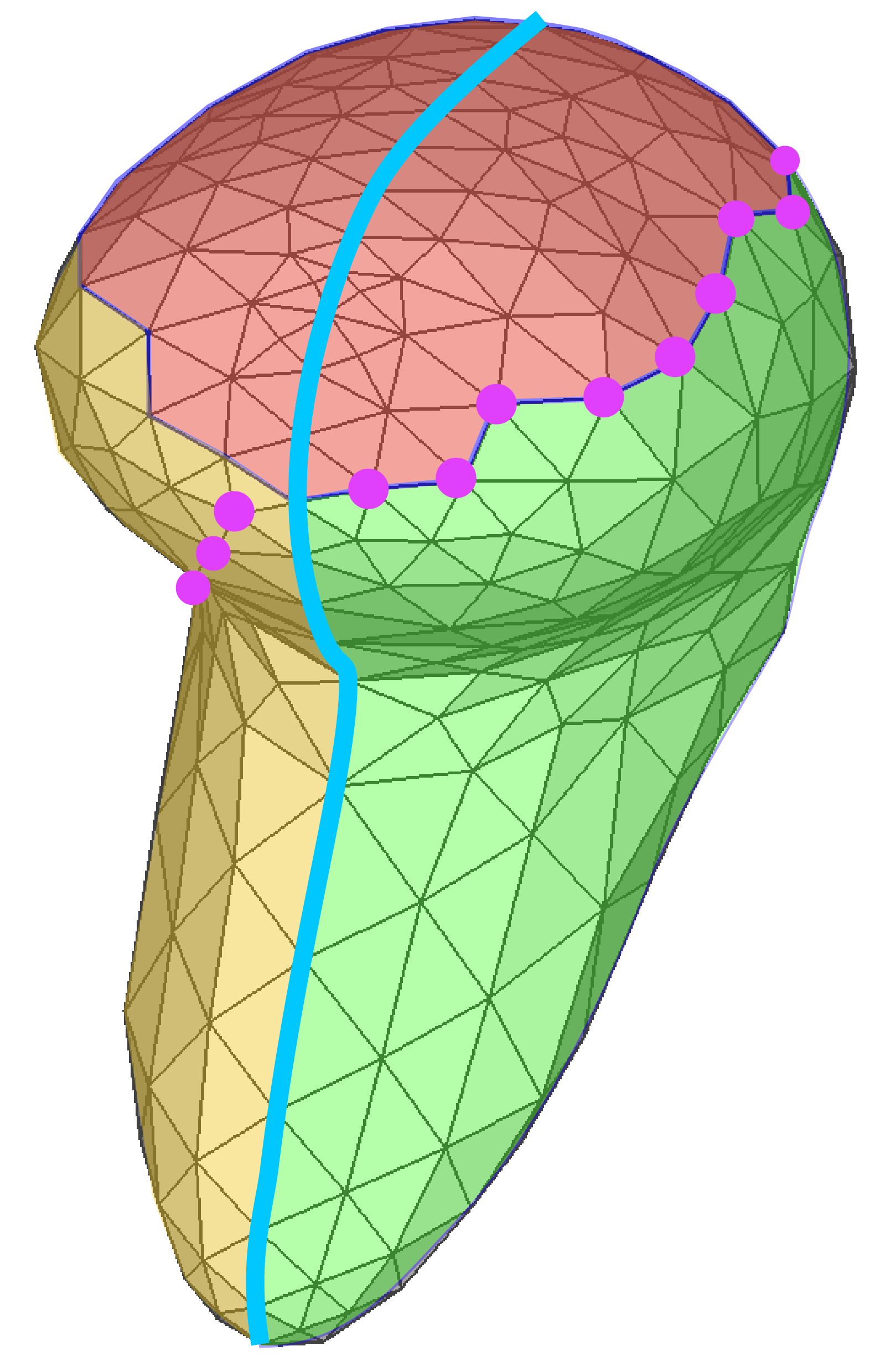}%
\label{fig:texture_visibility_b}}
\caption{Texture visibility. When there is an angle from the view plane (or the drawing plane) to the symmetry plane, there must be an invisible region (red region), even under our assumption of bilateral symmetry. We refer to this region as {\em ill-textured}. To handle this region, we propose a surface-based texture inpainting method to equip the final character model with a high-resolution texture.}
\label{fig:texture_visibility}
\end{figure}

We make use of the content in the reference drawing to texture the created shape. As the outline confines the region of the component in the drawing and bounds the 3D geometry surface as well, a natural starting point is simply to map the content in the confined region to the surface via orthogonal projection. This method produces a pleasing texture when viewed at the original viewpoint, but usually leads to distortion near the silhouette observed from novel views. To tackle this problem, we adopt the orthogonal projection as the initial state, and fix the texture coordinates on the silhouette and the midline (purple dots and the blue line in Fig.\ref{fig:texture_visibility_b}) as a boundary condition; we then solve for a harmonic mapping $h$ that maps $(x,y,z)$ from the surface mesh to the desired texture coordinates $(u,v) \in \mathbb{R}^2$:
\begin{equation}
\label{eq:harmonic_mapping}
\Delta u = 0,~~\Delta v = 0, ~~s.t. (u_i, v_i) = (\bar{u}_i, \bar{v}_i), i \in \mathcal{I},
\end{equation}
where $\Delta$ is the discrete Laplace operator computed on the surface mesh, $(u,v)$ are the texture coordinates of the surface, $(\bar{u}, \bar{v})$ are the texture coordinates of the drawing computed via orthogonal projection, and $\mathcal{I}$ are the indexing set of constraint vertices.

To avoid texture distortion near salient features (e.g. eyes), CreatureShop allows the user to add identified pairs of symmetric landmarks to the boundary conditions $\mathcal{I}$ to make sure that they look faithful from arbitrary views.

Using only a single reference drawing, it is impossible to faithfully recover the texture on the invisible side of the shape.
As we have assumed that all body parts are bilateral symmetric, we simply mirror the texture on the half side of the shape with larger visibility to fill the other side of the shape to cover as much region as possible.
This works perfectly only when the body part is presented from an exact side view; otherwise a small portion of the surface mesh is still ill-textured (the red region in Fig.\ref{fig:texture_visibility_b}). We resolve this final ill-textured region via a texture completion pass once all body parts are positioned correctly and merged together.

\section{Creating a character}

Our system infers a default depth for each created body part, so that they can be positioned for assembly; users may further adjust the depth when needed. Once all body parts are correctly positioned, the system merges them by performing a series of boolean union operations, smoothing the joining regions of the parts, and finally outputting a watertight character model with a full texture. The remaining issue of ill-textured regions is addressed automatically at this stage by a surface-based texture completion method.

In the following section, we detail the surface-based texture completion for the fully textured character creation. This is followed by brief discussions of other implementation details (i.e. default depth estimation and detection of ill-textured regions).

\subsection{Texture Inpainting over the Surface}

As explained in Section 4.3, incomplete textures are inevitable even when bilateral symmetry is leveraged for texturing surface geometry. We therefore aim to use content from the well-textured regions to inpaint any incompletely textured regions, with the eventual output being a seamless texture over the entire surface geometry of the character model after all constituent body parts are merged. Ill-textured regions may be disconnected and scattered over the surface. To avoid cluttered notation, we first consider the problem of surface-based inpainting in the case where there is only one connected ill-textured region. Our approach can then be easily generalized to handle multiple regions.

To avoid possible artifacts (including distortion or unexpected seams) due to one global planar parameterization, we perform the texture completion directly on the surface geometry by extending an image inpainting method \cite{Darabi2012} to surface meshes. We fulfill this goal by using the local Geodesic Polar Map (GPM) and Geodesic Polar Coordinates (GPC) \cite{Melvaer2012}. The GPM provides a local parameterization of a given region from curved meshes, while the GPC provides a local geodesic-preserving coordinate system in the parametric domain.

This extension from images to 3D surface meshes requires addressing the conflict between low-resolution meshes for efficient computation and high-resolution texture images for quality output. We note that while some previous work on surface-based texture synthesis and inpainting (e.g. \cite{Wei2001,Chen2012}) simply processes colors at mesh vertices, this approach is not sufficient for our needs as it fails to capture delicate patterns and details that may be present in the reference drawing.
To attain high-resolution textures with a low-resolution mesh, we therefore create a new texture image with sufficiently high resolution for the ill-textured target region under a new texture parameterization, which is built via a locally injective mapping approach \cite{Rabinovich2017}. Other parameterizations can be used as well; our results do not depend on the choice of parameterization, unless squeezed regions lead to insufficient resolution in areas when the parameterization is far from isometric.

We denote the ill-textured region as target region $\mathbf{T}$ and the complement of $\mathbf{T}$ on the surface as the source regions $\mathbf{S}$. Our task is to generate a consistent texture over $\mathbf{T}$ using the content from $\mathbf{S}$.
Two major steps are alternated in the general patch-match framework, namely a search step for finding similar patches from $\mathbf{S}$ and a color voting step to generate texture in $\mathbf{T}$ the concerned region.

Geodesic discs centered at mesh vertices are used as a substitute of the patches in image-based tasks.
Each geodesic disc $D(\mathbf{v}_i)$ covers a small number of faces surrounding its center vertex $\mathbf{v}_i$.
We build a local GPM for each vertex, which parameterizes covered faces in a geodesic preserving manner; we then construct a GPC within a geodesic disc of radius $r$. The collection of discs covering a portion of the target region is termed the {\em target set} $\mathcal{T}=\{D(\mathbf{v}_i)|D(\mathbf{v}_i) \cap \mathbf{T}\neq\emptyset\}$, and the collection of discs covering only the source region forms the {\em source set} $\mathcal{S}=\{D(\mathbf{v}_i)|D(\mathbf{v}_i) \subseteq \mathbf{S}\}$.

The geodesic disc is sampled uniformly in $\mathrm{GPC}_{\mathbf{v}_i}$ with respect to the polar coordinates defined on it.
The sample points on the disc can be represented as $\{(\theta_j,r_k)|1\leq j\leq n,1\leq k\leq m\}$ in the $\mathrm{GPC}_{\mathbf{v}_i}$. The source texture image has five channels: three color channels in LAB color space and two gradient channels of luminance. We operate on L*a*b color space following \cite{Darabi2012} and \cite{Barnes2009}, and employ the gradient following Darabi et al.'s conclusion that gradients are helpful when extending PatchMatch to preserve structural textures. Colors and gradients at sample points are concatenated (denoted as $\mathbf{I}(\theta_j,r_k)=(L,a,b,\lambda\nabla_xL,\lambda\nabla_yL)$; $\lambda=0.2$ is a constant balancing magnitudes between colors and gradients); we may therefore define the distance between two discs $D(\mathbf{v}_s)$ and $D(\mathbf{v}_t)$ as below:

\begin{equation}
d(D(\mathbf{v}_s),D(\mathbf{v}_t))=\underset{0\leq \ell \leq n-1}{\min}\sum_{j,k}\|\mathbf{I}_s(\theta_{j}+\ell\Delta\theta,r_k)-\mathbf{I}_t(\theta_j,r_k)\|.
\end{equation}
where $\{\theta_j\}$ are uniformly sampled in a counterclockwise direction with a step of $\Delta\theta={2\pi}/{n}$,
The optimal matching angle between the two geodesic fans can be represented as $\alpha=\tilde{\ell}\Delta\theta$.

To efficiently find the optimal matching disc in $\mathcal{S}$ for every disc in $\mathcal{T}$  using the disc similarity distance defined above, we use a search method similar to \cite{Chen2012} which extends the well-known image patch-based search method PatchMatch to meshes. We call the mapping from a disc to its optimal matching disc the {\em Nearest-Neighbour Field (NNF)}.

\begin{figure}[htbp]
\centering
\includegraphics[width=0.8\linewidth]{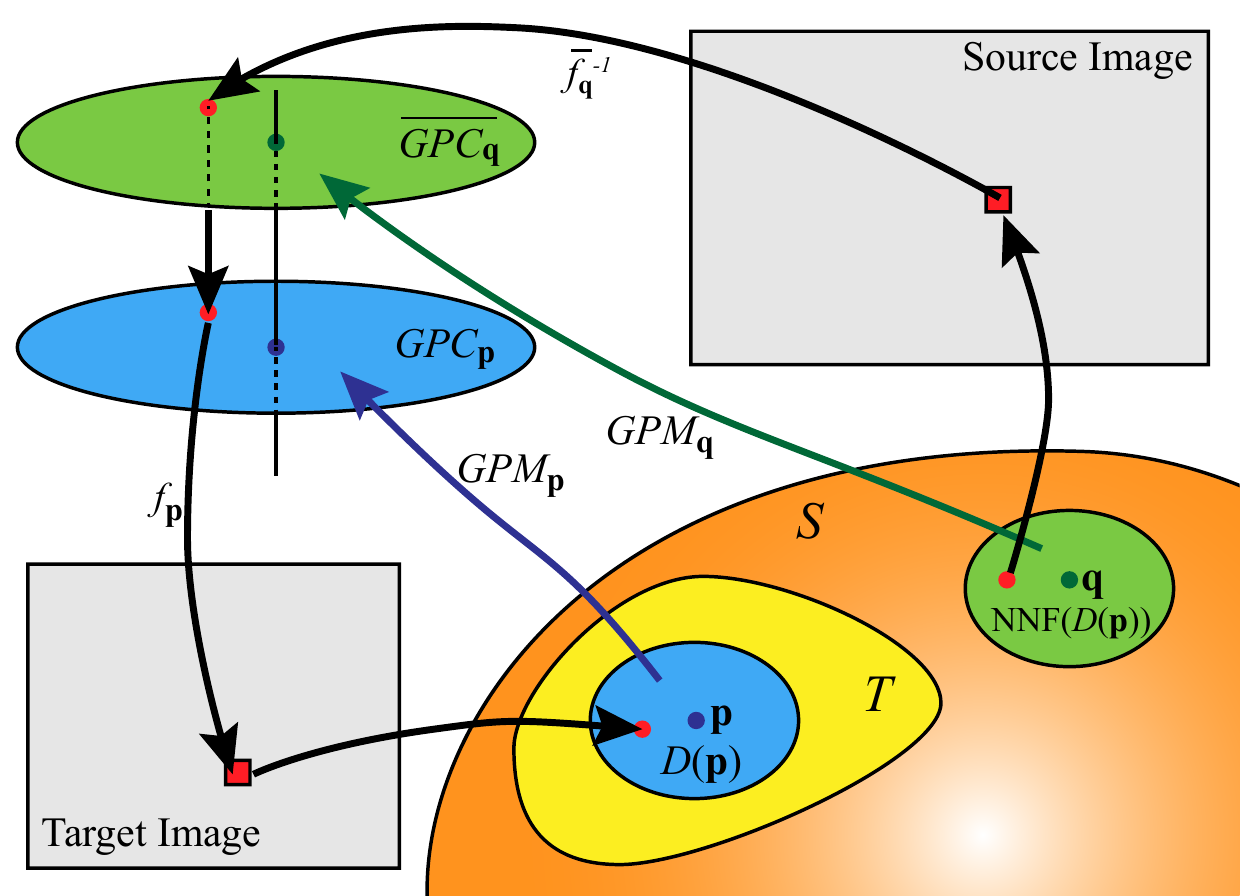}
\caption{Voting procedure in our surface inpainting method. For an unknown $\mathbf{p}$ in the target surface region $T$, the colors in its disc $D(\mathbf{p})$ are voted on by its optimal matching disc $D(\mathbf{q})=\mathrm{NNF}(D(\mathbf{p}))$.}
\label{fig:voting}
\end{figure}

After finding NNFs in the search phase, a new texture is generated via voting; the colors and gradients of each pixel in the target texture image is voted on by all the discs that cover it in the texture domain. Backtracking the colors (and gradients) of a pixel in the target image from the source image is not straightforward, requiring passing through two GPCs defined on target and source geodesic discs, respectively; see Fig.\ref{fig:voting}. In this process, the pixel of interest needs to be located in each GPC by interpolating the texture coordinate in the corresponding parameter domain.
For a median-resolution target texture image with 500K pixels, this location query could be computationally prohibitive.
To accelerate this procedure, we propose the following approach that is well suited to parallel processing using GPU shader pipelines.

As illustrated in Fig.\ref{fig:voting}, for each disc $D(\mathbf{p})$ from target set $\mathcal{T}$, we have its NNF $D(\mathbf{q})$ from source set $\mathcal{S}$ obtained in the searching step.
First, we rotate the $\mathrm{GPC}_{\mathbf{q}}$ clockwise by the optimal matching angle $\alpha$ about its center $\mathrm{GPM}_{\mathbf{q}}(\mathbf{q})$ to generate $\overline{\mathrm{GPC}_{\mathbf{q}}}$. A new texture mapping $\overline{f_\mathbf{q}}$ can then be obtained which maps $\overline{\mathrm{GPC}_{\mathbf{q}}}$ to the source image domain.
Second, we rasterize the textured $\overline{\mathrm{GPC}_{\mathbf{q}}}$ to a high-resolution intermediate image $\tilde{\mathbf{I}}$ and apply to $\tilde{\mathbf{I}}$ a Gaussian falloff filter centered at $\mathrm{GPM}_{\mathbf{q}}(\mathbf{q})$, which can be finished via GPU pipeline.
Third, using the intermediate image $\tilde{\mathbf{I}}$ as the texture of $\mathrm{GPC}_{\mathbf{p}}$, we map $\tilde{\mathbf{I}}$ to the target image via the texture mapping $f_\mathbf{p}$.
Each pixel in the target image contains (three) color channels, (two) gradient channels and a counting channel, where the color and gradient channels accumulate relevant values voted from $\tilde{\mathbf{I}}$, and the counting channel keeps record on the number of votes. This phase can be finished by alpha blending using GPU.
Finally, we compute the colors and gradients at each pixel in the target voting image $\mathbf{I}_\text{vote}$ by dividing the accumulated colors and gradients at this pixel by the number of covering discs. To finalize the voting results, the new color $\mathbf{I}_\text{new}$ can then be computed by minimizing the energy 

\begin{equation}
E=\sum (\mathbf{I}_\text{new} - \mathbf{I}_\text{vote})^2 + \lambda \|\nabla \mathbf{I}_\text{new} - \nabla \mathbf{I}_\text{vote}\|^2
\end{equation}
to correctly blend colors and gradients; this is a discrete screened Poisson equation that can be solved efficiently \cite{Darabi2012}.

After each voting phase, the geodesic fans in $\mathcal{T}$ and the optimal disc distances $d(D(\mathbf{p}))$ are updated. We compute the total energy $E=\sum_{D(\mathbf{p})\in\mathcal{T}}d(D(\mathbf{p}))$, and compare it to the previous pass.
We iterate the search and vote steps until the difference of total energies between two sequential passes is less than some threshold (we use $0.01$ in our implementation). We then rebuild all the discs using a smaller disc radius $\tilde{r}$, update set $\mathcal{S}$ and $\mathcal{T}$, and repeat the search and voting phases. This process terminates once the radius reaches the pre-defined threshold.
The pseudo-code of our algorithm is shown in Alg.\ref{alg_inpainting}.
\IncMargin{1em}
\begin{algorithm}
\SetKwInOut{Input}{Input}\SetKwInOut{Output}{Output}
\Input{Well textured source regions $S$ and target regions $T$ to be filled}
\Output{Compatible textured entire surface}
\BlankLine
Initialize $T$ by solving Poisson equation with boundary condition on $\partial T$\;
Assign a large fan distance $d(\mathbf{p})$ to each disc $D(\mathbf{p})\in\mathcal{T}$\;
\For{scale $r$ from $r_{max}\rightarrow r_{min}$ with step size $r_s$}{
\While{True}{
Compose geodesic discs set $\mathcal{S}$ and $\mathcal{T}$ with radius $r$\;
Randomly select a disc $D(\mathbf{p}_0)\in\mathcal{T}$\;
\ForEach{$\mathbf{p}$ visited in breadth-first traversal order started from $\mathbf{p}_0$ on the mesh and $D(\mathbf{p})\in\mathcal{T}$}{
Collect all adjacent points that have been traversed in $\{a_i(\mathbf{p})\}$\;
\ForEach{$\mathbf{q}$ that is the center of $\mathrm{NNF}(D(a_i(\mathbf{p})))$}{
Randomly select a disc $D(\tilde{\mathbf{q}})\in\mathcal{S}$\;
$d(\mathbf{p})\leftarrow \min \{d(\mathbf{p}),dist(D(\mathbf{p}),D(\mathbf{q})),$
$dist(D(\mathbf{p}),D(\tilde{\mathbf{q}}))\}$\;
Update $\mathrm{NNF}(\mathbf{p})$\;
}
}
Vote(NNF)\;
Recompute all the geodesic fans and $d(\mathbf{p})$\;
\lIf{Total energy unchanged}{break}
}
}
\caption{Texture inpainting over the surface}
\label{alg_inpainting}
\end{algorithm}\DecMargin{1em}

\subsection{Assembling and unifying body parts}

To facilitate the positioning of body parts created, we design two ways for users to indicate whether body parts are paired (or intrinsically symmetric) that reside at both sides of the existing parts.
If a body part is created alone, then its initial depth and symmetry plane are inferred using a \emph{central positioning} scheme; otherwise, we use a \emph{biased positioning} scheme to estimate depths and symmetry planes for both body parts. In the latter case, the textures of the two body parts are enforced to be as similar as possible as well via a further step of texture transfer from one to the other.

\paragraph{Central Positioning}
When a new individual component $\mathcal{M}_i$ is created and its outline $o_i$ overlaps those of other existing parts, we find the part $\mathcal{M}_j$ whose outline has the largest overlapped region with $o_i$.
Reusing the symmetry plane $\pi_j$ of $\mathcal{M}_j$, we can create the base shape of $\mathcal{M}_i$ automatically.
Sometimes this rule will lead to a result where $\mathcal{M}_i$ and $\mathcal{M}_j$ have no actual intersection in 3D.
In such a situation, we shoot a ray passing through the center of overlapped region of the two outlines and find first two intersection points with $\mathcal{M}_j$ in 3D space.
We then compute their average depth $d$ and set the symmetry plane of $\mathcal{M}_i$ to $\pi_i:z=d$ to create a base shape.

\paragraph{Biased positioning}
When two intrinsically symmetric components $\mathcal{M}_i$ and $\mathcal{M}_j$ are created together, their outlines should overlap with a common parent part $\mathcal{M}_k$.
We shoot a ray passing through the center of the overlapped region of outlines $o_i$ and $o_k$ and find the first intersection point $\mathbf{p}_i(x_i,y_i,z_i)$ with $\mathcal{M}_k$. Set $\pi_i:a_k(x-x_i)+b_k(y-y_i)+c_k(z-z_i)=0$ to be the symmetry plane of $\mathcal{M}_i$, where $(a_k,b_k,c_k)$ is the normal of plane $\pi_k$. Then the base shape of $\mathcal{M}_i$ can be generated. The same way can be adopted again to assemble $\mathcal{M}_j$ except that the intersection point $\mathbf{p}_j$ should be found on the opposite side of $\pi_k$ against $\mathbf{p}_i$.

This system provides users with only an initial guess which can be further edited easily. Users can adjust the depth of each component to change the assembly relation to its connected components according to user preference.
Users may pose the shape by orienting its symmetry plane as described in section 4.2.3. The rotation center in this case is the center of junctions, so that connectivity between the new shape and its parent shape is ensured.

\begin{figure}[htbp]
\centering
\subfloat[]{\includegraphics[height=0.3\linewidth]{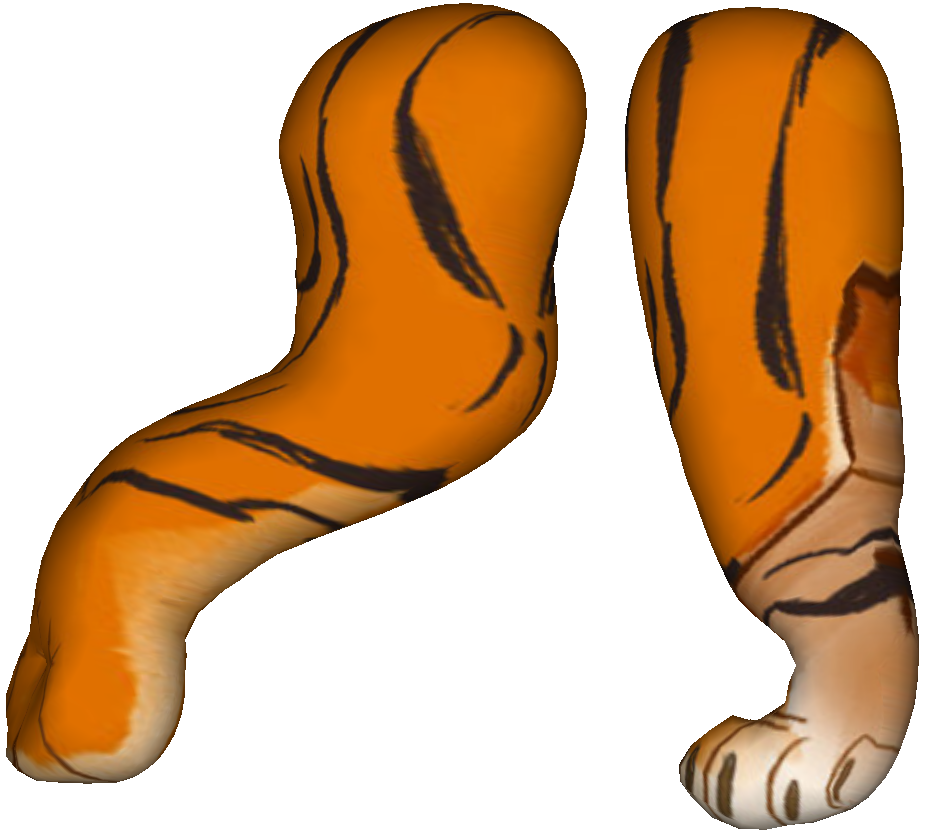}%
\label{fig:texture_transfer_a}}
\hfil
\subfloat[]{\includegraphics[height=0.3\linewidth]{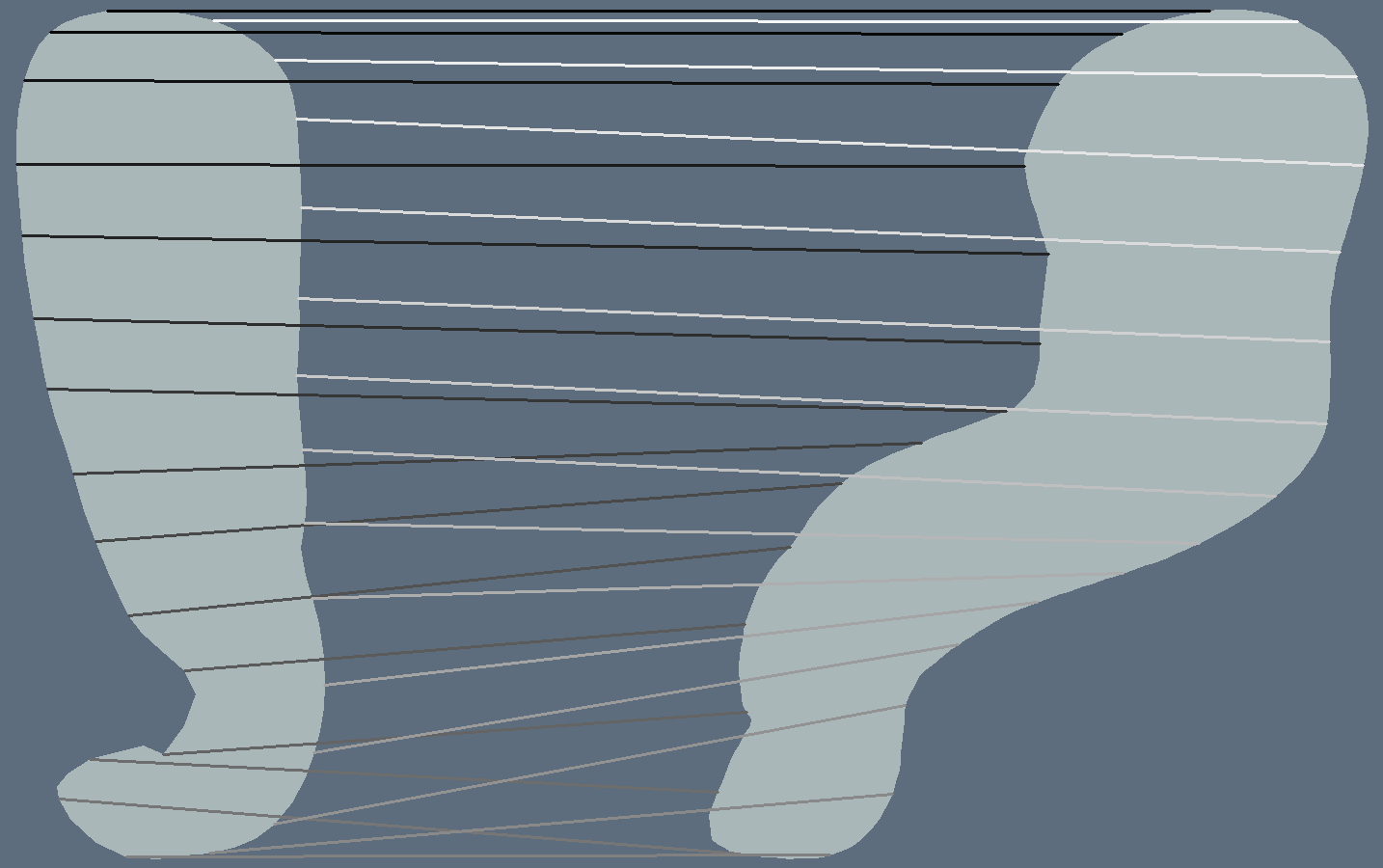}%
\label{fig:texture_transfer_b}}\\
\subfloat[]{\includegraphics[height=0.45\linewidth]{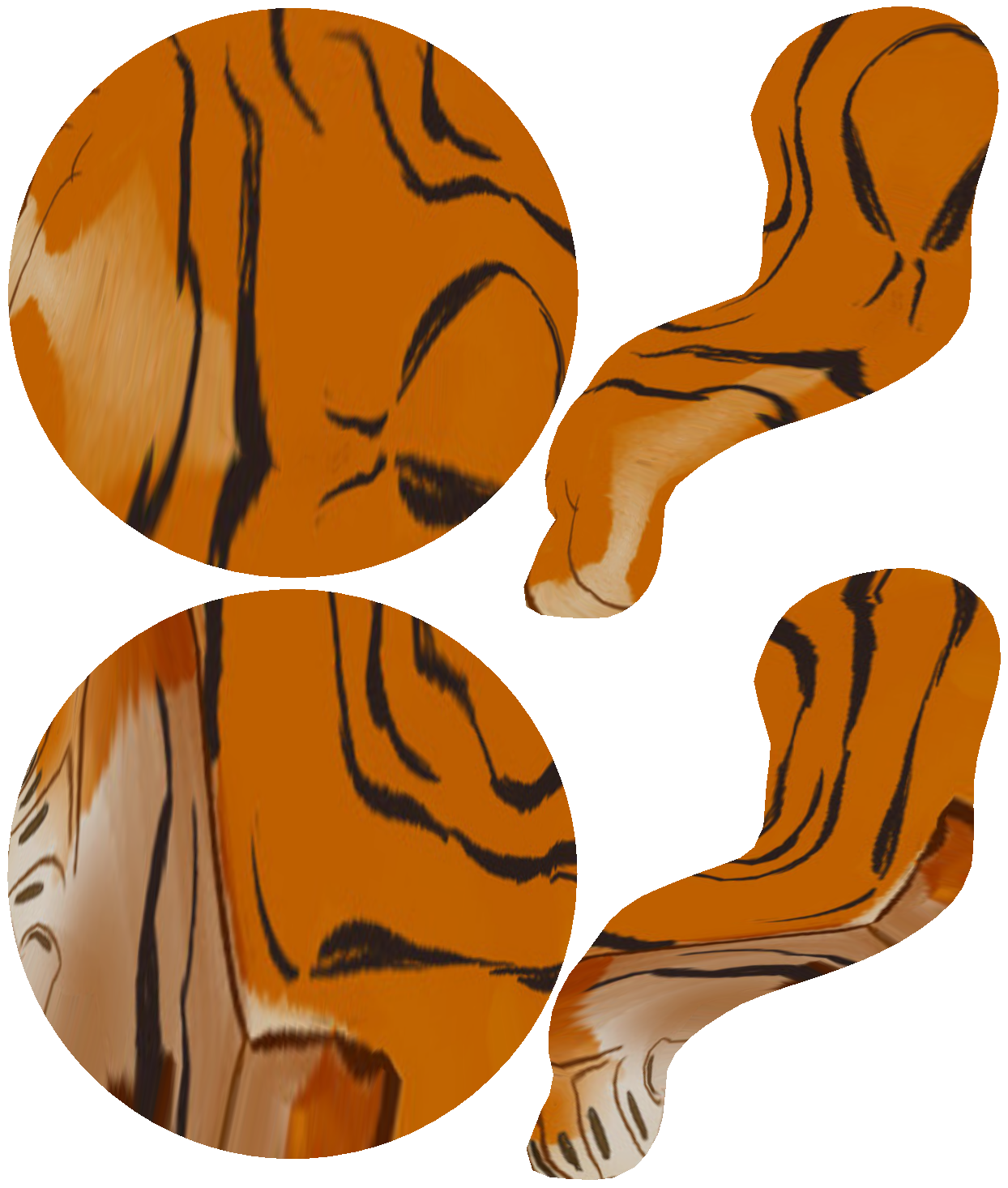}%
\label{fig:texture_transfer_c}}
\hfil
\subfloat[]{\includegraphics[height=0.45\linewidth]{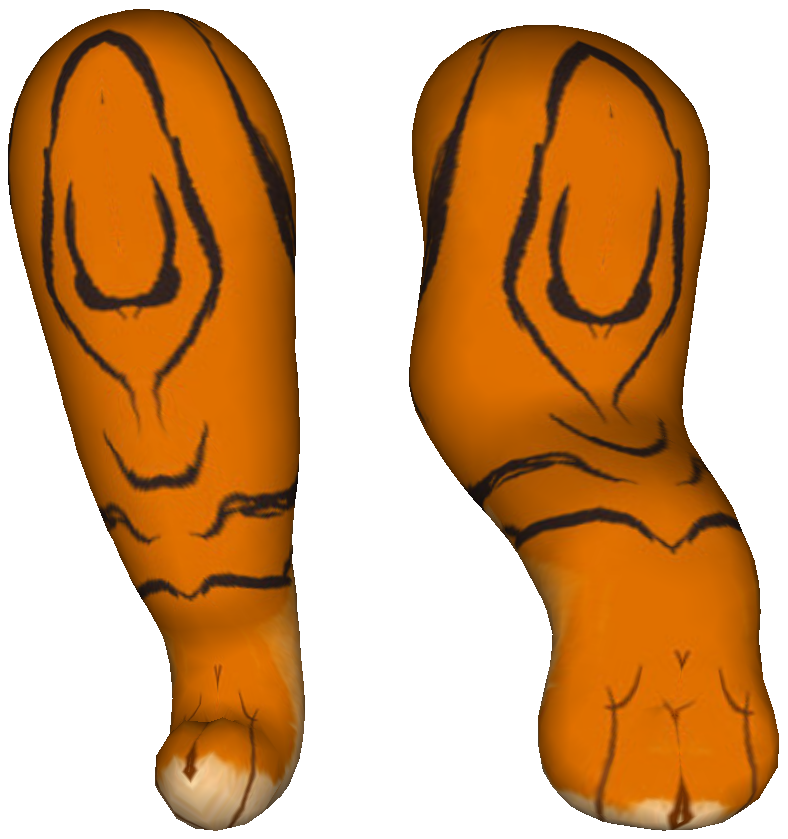}%
\label{fig:texture_transfer_d}}
\caption{Texture transfer between two intrinsically symmetric body parts. (a) illustration of two body parts, and the right one has the wrong texture; (b) pointwise correspondences achieved by IDSC method \cite{Ling2007}; (c) the proxy circle domains (left) and the planar parameterization generated by \cite{Rabinovich2017} (right); (d) the transferred texture from the left one to the right one.}
\label{fig:texture_transfer}
\end{figure}

When two body parts $\mathcal{M}_i$ and $\mathcal{M}_j$ are created as intrinsically symmetric, the partially occluded one $\mathcal{M}_j$ usually has the wrong texture (see Fig.\ref{fig:texture_transfer_a}), since its texture directly comes from its projected region in the image. We therefore must copy the texture from the occluded-free body part $\mathcal{M}_i$ to $\mathcal{M}_j$. We compute a set of pointwise correspondences between the symmetry curves $M_i$ and $M_j$ via inner-distance shape context \cite{Ling2007} (Fig.\ref{fig:texture_transfer_b}). By confining the matching results to have at least one pair that maps from one junction region to the other, we can get rid of the upside down mismatch between two tube-like limbs. We also enable users to specify several key correspondence pairs if the poses of two parts are largely different.
We then compute planar parameterizations of $\mathcal{M}_i$ and $\mathcal{M}_j$, both bounded by the symmetry curves. To avoid severe flips in such a parameterization, we first transform the entire region to a proxy convex planar region (a circle in our implementation) using a Harmonic mapping (see Fig.\ref{fig:texture_transfer_c}); we then deform the boundary of each circle to the shape of the symmetry curve $M_i$, which is based on the correspondences built in the first step for the occluded one, and compute the positions of inner points using \cite{Rabinovich2017}, see Fig.\ref{fig:texture_transfer_c}.
Finally, for arbitrary position on the occluded body part $\mathcal{M}_j$, we can interpolate a position on $\mathcal{M}_i$ via this parameter domain. Hence, the texture coordinates from $\mathcal{M}_i$ can be transferred to $\mathcal{M}_j$, see Fig.\ref{fig:texture_transfer_d}.

We then unify the geometries and their texture data by performing boolean union operations among all connected parts. The merged geometry is smoothed by Laplacian smoothing to avoid unnatural junction, and we interpolate texture coordinates of vertices in the new mesh by querying the closest point in the original mesh. Texture may have discontinuities around junctions especially on occluded sides; to overcome such issues we compute harmonic mappings in all junction regions to stitch their texture coordinates. Texture coordinates continuity will result in seamless textures, since the original drawing image is used as the sole common texture image for all parts before the inpainting phase.

After merging the geometries as well as stitching the textures, we obtain a unified model with watertight geometry but implausible textures in regions that were occluded or not present in the original drawings. To address this problem, we detect all ill-textured regions and furnish them with textures using a surface-based texture inpainting technique as described in Section 5.1. Detecting ill-textured regions is made non-trivial, owing to the harmonic mapping applied to minimize texture distortion.
Since we leverage mirror symmetry to texture individual body parts, we determine the so-called positive facets in the mesh which reside at the positive side of the symmetry plane, and then determine visibility of the negative facets based on their symmetric correspondences, reducing the complexity of the problem. 
As harmonic maps are used to transfer texture from the drawing to the surface, we rely on the UV texture coordinates of each facet, instead of its spatial coordinates, to determine its visibility over the texture. In particular, when a texture coordinate may be referenced by several facets, we set only the facet which is least distant from the original viewing position to visible; all others are considered invisible as they are occluded by the visible facet.
\section{Experimental Results and Validation}
\begin{figure*}[htbp]
  \centering
  \includegraphics[width=0.99\textwidth]{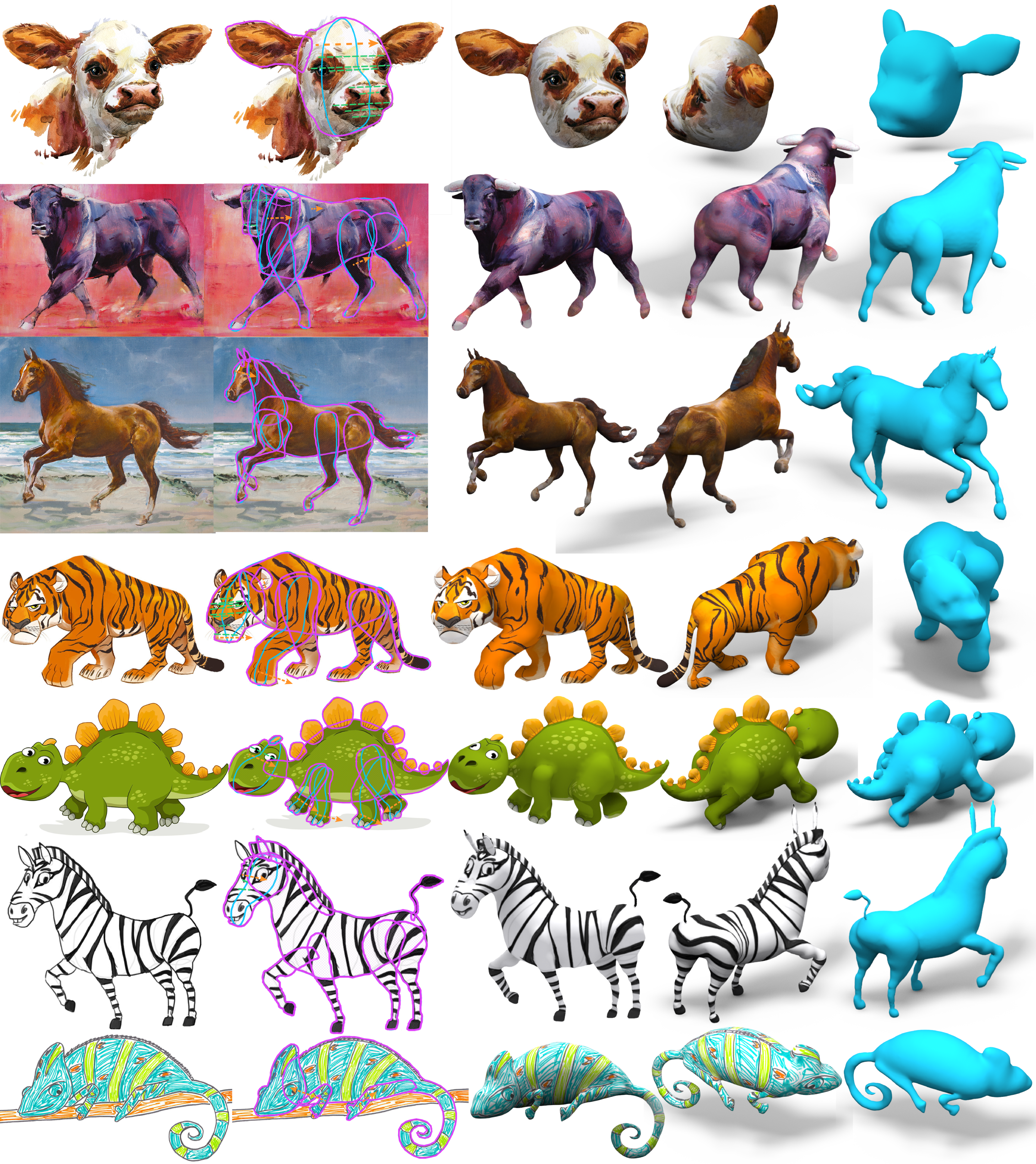}
  \caption{Left to right: input drawings, user's markers with CreatureShop, output models and textures presented from the original and novel viewpoints. Original images in Row 1, 3, 7 are purchased under license from \url{https://www.dreamstime.com}. Levantado by Mark Adlington in Row 2 licensed for journal publication use from \url{https://www.bridgemanimages.com}. Dinosaur in Row 5 is purchased under standard licensing terms from \url{https://www.canstockphoto.com}. Zebra in Row 6 obtained under CC-by-SA license from \url{https://www.wikihow.com}.}\label{fig:result_gallery}
\end{figure*}

We demonstrate the versatility of our method by exhibiting a dozen character models with full textures that are created using CreatureShop. Additionally, we validate our algorithms and design choices by comparison against ground truth models; comparisons to previous work; a small-scale, informal user study; and by 3D printing toy creature models made using our system.

If not explicitly marked by annotations, body parts shown in this section are made without thickness/depth adjustments.
We find that torsos and limbs sometimes need additional rotation when they do not share the same orientation as heads; ears and tails always need thickness adjustment.

\paragraph{Results} The gallery in Fig.\ref{fig:result_gallery} presents several 3D textured models created by CreatureShop, including characters of different types from reference drawings realized in different media or styles.
In each row, we show (from left to right): the input drawing; the input drawing with user annotations; the output of CreatureShop from the original view; the output of CreatureShop from a novel view; and the unshaded surface geometry. User annotations are colored as follows throughout the paper: outlines in purple, midlines in blue, user's specified rotation axes in orange, and symmetric features in green.

As the heads of all creatures have plenty of features in the reference drawings, we use the outline and the symmetry plane rotation for each as obligatory inputs when shaping them. We incrementally edit midlines or add symmetric texture features as optional operations until satisfying details are added to the shape.
Torsos and limbs usually lack symmetric landmarks, and thus are shaped purely using the outlines. The orientation can be evaluated by inheriting from the adjacent part automatically. If the user's desired orientation is not achieved, users can further drag rotation instruction segments to change the orientation (orange lines in our gallery show such additional interaction by users).
In our user interface, outlines are furnished by segmentation from character contours; symmetric texture features can be indicated from visual cues; midlines can be edited by picking the control points on midline and dragging to its target position in drawings. When symmetric features are absent, rotation instruction segments are dragged by users to obtain an oblique symmetry plane according to its direction and length (following the rules described in 4.2).
Modeling oblique parts with symmetry features can generate satisfying results directly without further adjusting the orientation of the symmetry plane in all of our results.

The rotation preview can give users a real-time feedback in the modeling process.
Some objects with thinner geometry (e.g. ears) can be shaped by adjusting a slider bar in our user interface which controls $\alpha_c$ in equation.\ref{eq:position_solver} in the modeling section.
These user's operations yield compelling geometries as shown in the output models from both original and novel views, supporting our design choices and the claim that quality geometry can be created easily with the designed set of user annotations.

\begin{figure}[htbp]
\centering
\subfloat[Before inpainting]{\includegraphics[width=0.45\linewidth]{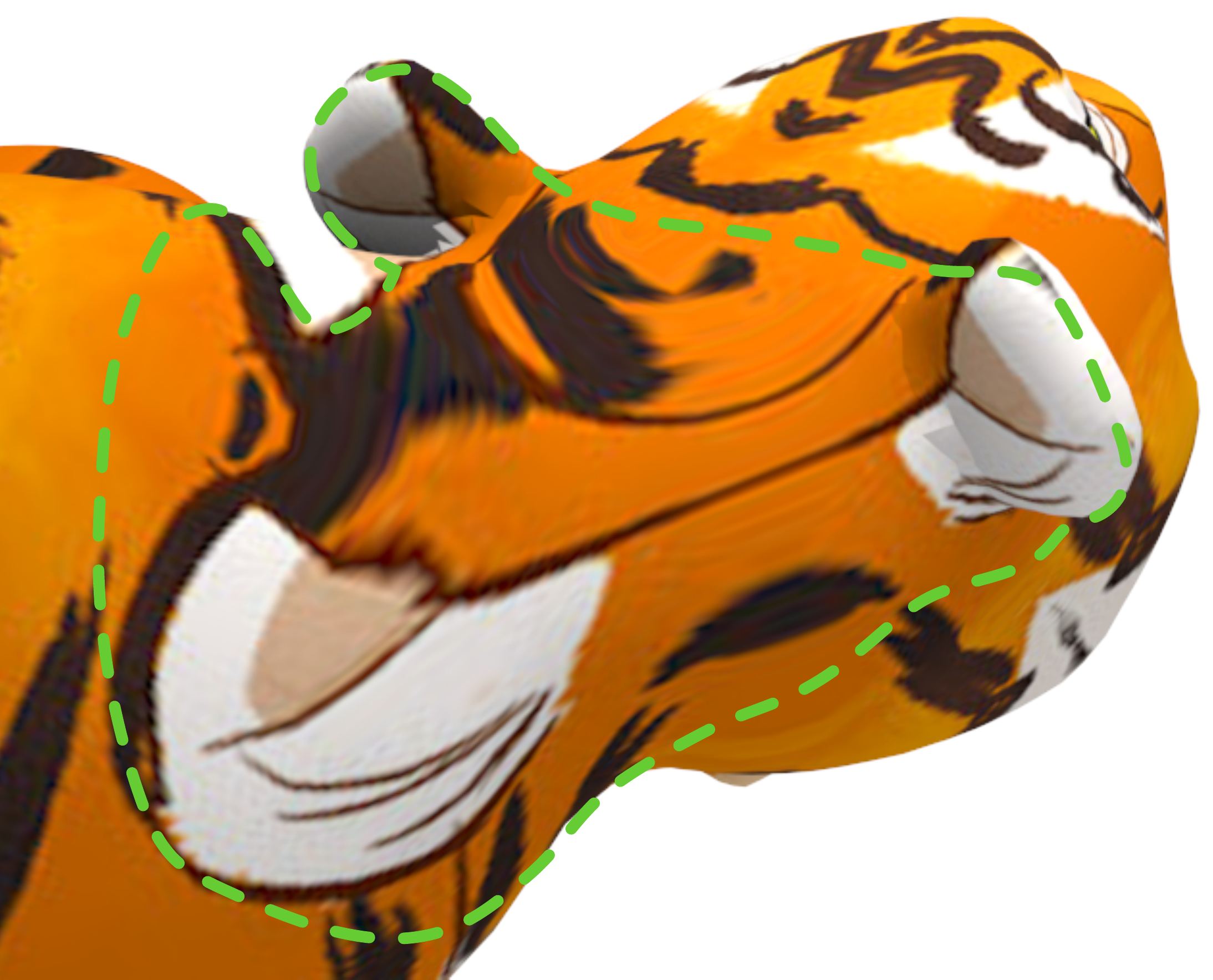}%
\label{fig:texture_result_a}}
\hfil
\subfloat[After inpainting]{\includegraphics[width=0.45\linewidth]{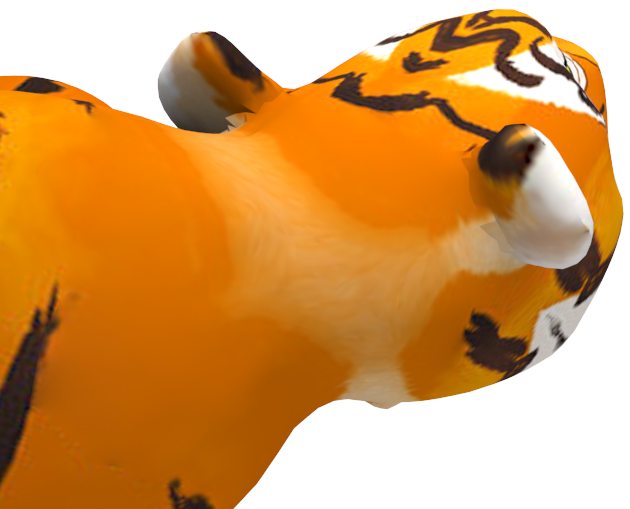}%
\label{fig:texture_result_b}}
\caption{Texture inpainting result. The occluded region (marked as the inner region within the green dash line) is completed by reasonable textures.}
\label{fig:texture_result}
\end{figure}

We also exhibit several character models produced from drawings with complicated textures. We are able to convincingly synthesize seam-free textures for the entire model. Fig.\ref{fig:texture_result} shows details of our texture inpainting result on the tiger model with zoom-in views. No visible seam can be seen on the boundary of the ill-textured region and the colors in the region are seamless and correctly blended across the entire model. The result appearance is natural and reasonable, considering they were generated from only a single view drawing with no additional human editing or repainting.

\paragraph{Runtime Efficiency.} The efficiency of our inpainting method is related to the number of vertices and the size of new texture images. In our implementation, the vertices number of models ranges from 10k to 20k. The new texture images are set to $400\times 400$. It costs less than 2 minutes each model to complete the texture inpainting process in our experiments.

\begin{figure}[ht]
\centering
\subfloat[Input image\label{fig:validation_a}]{\includegraphics[width=0.28\columnwidth]{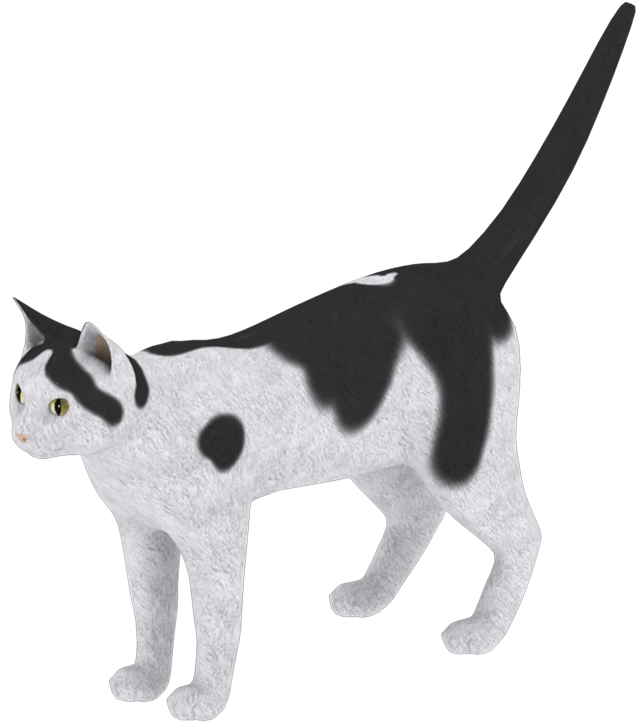}}
\hfil
\subfloat[Result created by CreatureShop\label{fig:validation_b}]{\includegraphics[width=0.28\columnwidth]{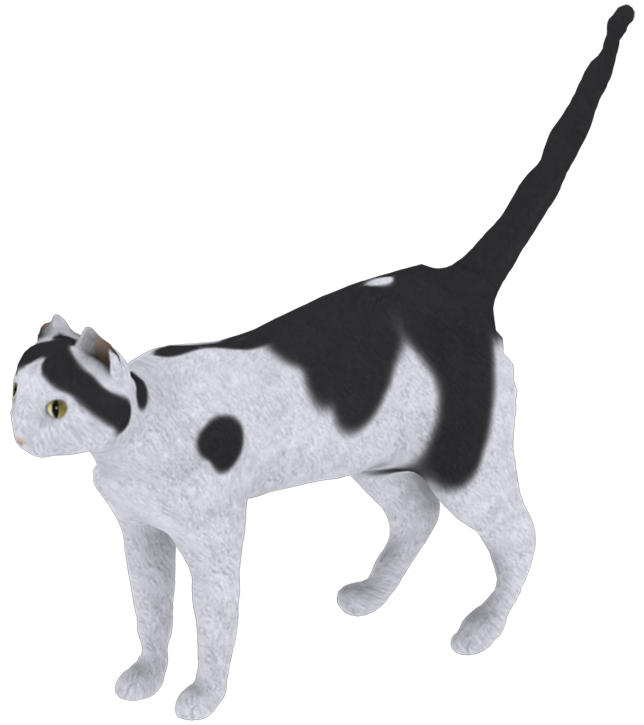}}
\hfil
\subfloat[Visualization of the per-vertex error  \label{fig:validation_c}]{\includegraphics[width=0.28\columnwidth]{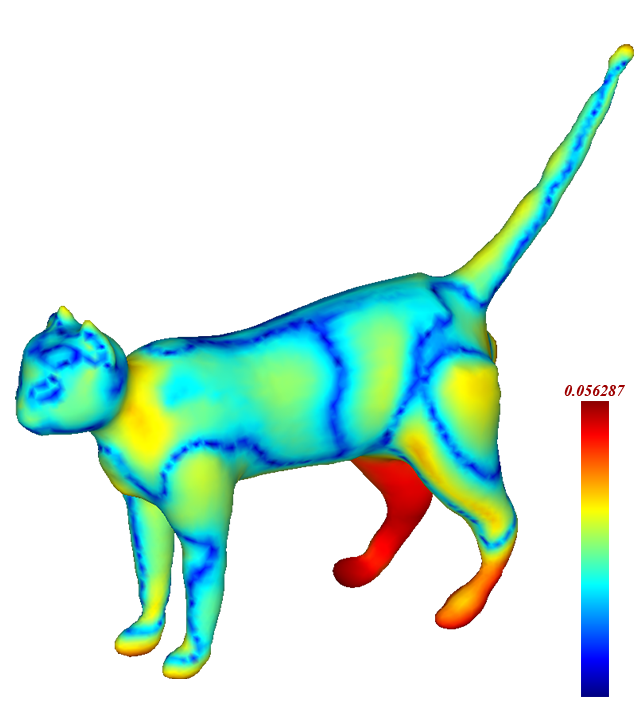}}
\caption{Validation of our method through comparison between a ground-truth model (left) and a reconstructed model using CreatureShop.}
\label{fig:validation}
\end{figure}

\subsection{Validation}
We quantitatively validate our geometry modeling method by comparing the created character to a ground-truth model, shown in Fig.\ref{fig:validation}. The ground truth is a given 3D model and the reference image is rendered from a known view angle.
The model was created with CreatureShop using the following order of operations: segmentation to body parts, modeling separate (pick eyes as symmetric features to compute symmetry plane of the head and other parts inherit the head orientation), and finally blending together and inpainting ill-textured regions.
Geometry errors are depicted in the color map with red indicating larger errors. The mean error is $0.8262\%$ with regard to the bounding box diagonal of the reference model and the max error is $5.6287\%$. One of the hindmost legs of the cat deviates notably from the ground truth, which is because the symmetry plane of the torso is not evaluated accurately. The ground-truth normal of the symmetry plane is $(0.695,-0.228,0.683)$, while the estimated normal is $(0.617,-0.248,0.746)$, the angle between them is $0.0946$ in radian (no more than 6 in degree). We consider this deviation tolerable for an interactive modeling tool that converts 2D bitmap drawing to 3D models.

\begin{figure}[ht]
  \centering
  \includegraphics[width=0.95\columnwidth]{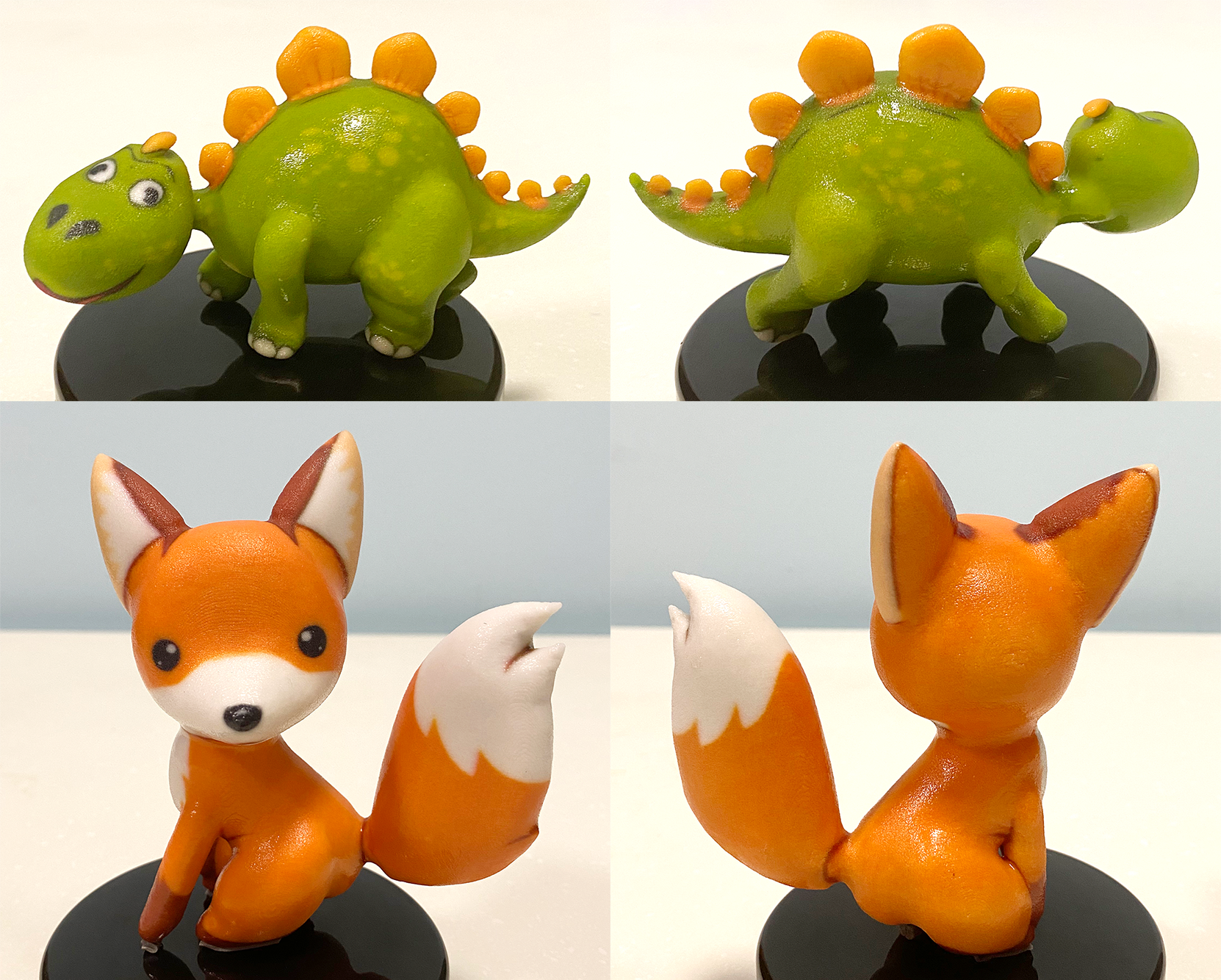}
  \caption{3D printed toy models using the textured character models made by CreatureShop.}\label{fig::fabrication}
\end{figure}

\subsection{Printed Toys}
As the created models by CreatureShop are watertight and readily for fabrication,
we also present some of the 3D printed version of character models by our system in Fig.\ref{fig::fabrication}. The models are printed using a ProJet CJP 660Pro with VisiJet PXL material.

\subsection{User Studies}
\begin{figure*}[htb]
  \centering
  \includegraphics[width=0.96\textwidth]{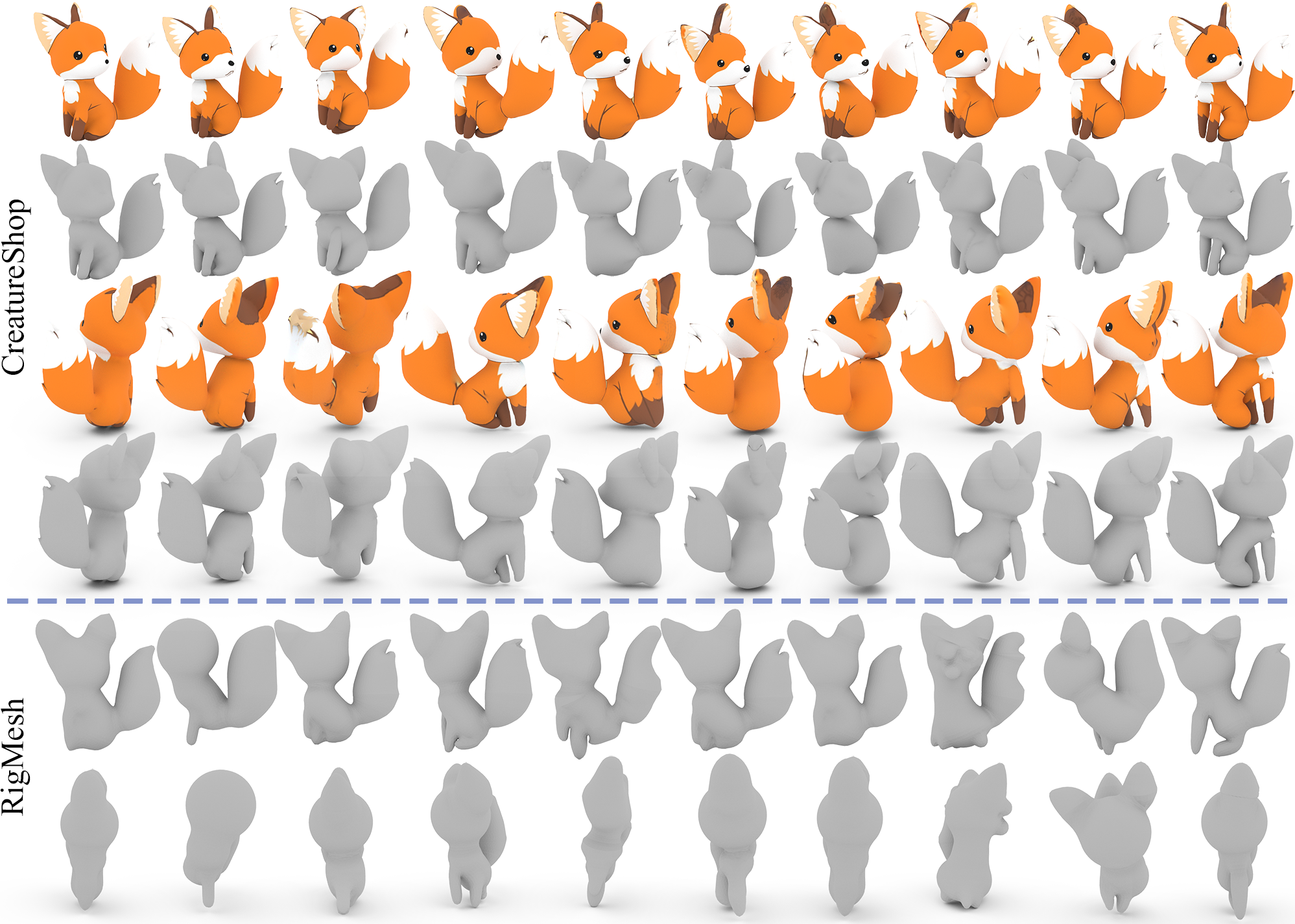}
  \caption{Character models created by our users in the user study. The top 4 rows were created with CreatureShop, showing the original and novel views with and without textures. The bottom 2 rows were created with RigMesh, showing the original view and a side view.}\label{fig:user_study}
\end{figure*}
The UI design is examined through a small-scale, informal user study to collect feedback. This user study consisted of 10 amateur users with few prior modeling experience, 4 female and 6 male. All the participants had no artistic inclinations; none of our users were professionals. Subjects were asked to create, with CreatureShop, a single part with symmetric landmark pairs and an entire character as shown in Fig.\ref{fig:modeling_with_detail_a} and Fig.\ref{fig:teaser}, respectively.

A 20-min tutorial was provided prior to the modeling test, which introduced the four annotations and basic operations in CreatureShop. The subjects were given 20 minutes to get hands-on experience with the system, and they all agreed that specifying symmetric features and editing midlines were easy tasks for them. During the tutorial, they only practiced basic operations using reference images that are different from target images.

After the tutorial, they were given two modeling tasks as described above. From the result, we can see the time for modeling the single body part (the gorilla's head shown in Fig.\ref{fig:modeling_with_detail_a}) ranges from 2 to 4 minutes.
Modeling the fox character (shown in Fig.\ref{fig:teaser}) required more time, ranging from 8 to 14 mins. We consider this reasonable because the fox character has 7 body parts in total. Most of the body parts are created with outlines and rotated symmetry planes.

From the interview with the subjects after the modeling tests, we were told that all the interactions are easily mastered and both the resulting shape and texture are satisfactory.

To compare our system against prior art, we asked users to model the fox character using RigMesh \cite{Borosan2012} as well. Similarly, we provided a 20-min tutorial and a 20-min practice for RigMesh. Users were then required to create a 3D model based on a reference image that was imported to RigMesh as the background image within 30 mins. We then compared the usability of CreatureShop and RigMesh by using the System Usability Scale (SUS) \cite{Brooke1996SUS}. The SUS score of CreatureShop is 72.5; in contrast, RigMesh is 47.25. The significance of the comparison was verified by a Wilcoxon Signed-Rank test ($p=0.0215$). We conclude that for the task of creating 3D models from single images, CreatureShop is more user-friendly for novice users than RigMesh. We show the user models created during this comparison in Fig. \ref{fig:user_study}; we observe that the models created by CreatureShop are more plausible than those created by RigMesh, especially when seen from a novel view.

\subsection{Comparison to Prior Art} Our results are compared to several state-of-the-art methods for character modeling.

\begin{figure}[htbp]
\centering
\subfloat[]{\includegraphics[width=0.24\linewidth]{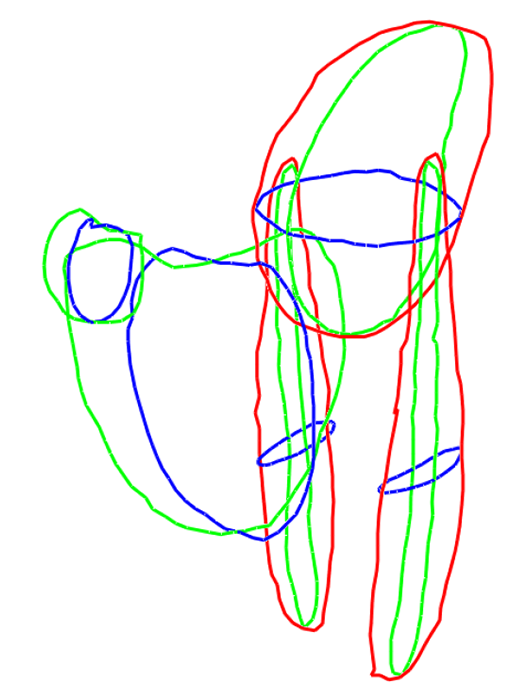}%
\label{fig:comparison_andre_a}}
\hfil
\subfloat[]{\includegraphics[width=0.24\linewidth]{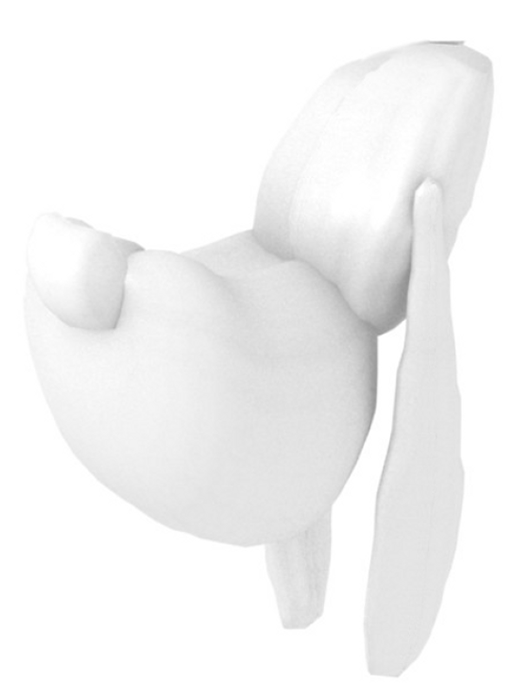}%
\label{fig:comparison_andre_b}}
\hfil
\subfloat[]{\includegraphics[width=0.24\linewidth]{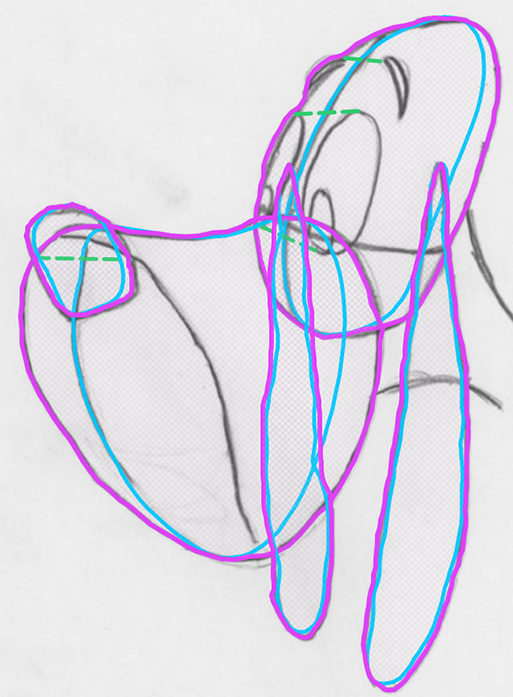}%
\label{fig:comparison_andre_c}}
\hfil
\subfloat[]{\includegraphics[width=0.24\linewidth]{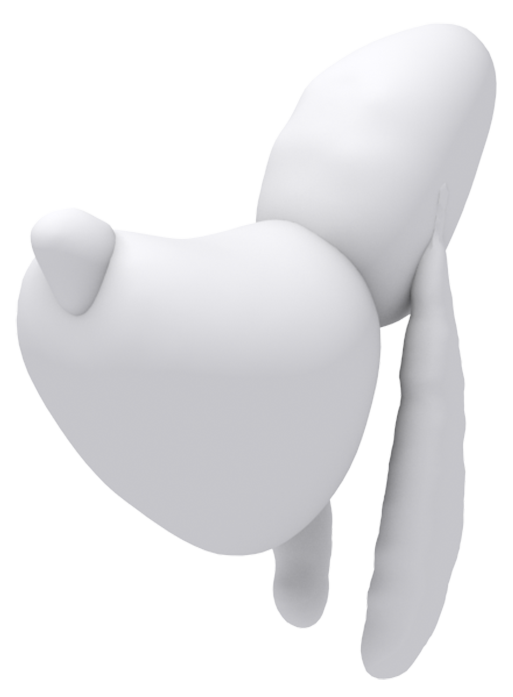}%
\label{fig:comparison_andre_d}}
\caption{Comparison with the interactive modeling method in Andre \& Saito \cite{Andre2011}. Their user's annotations (a) are more complex than ours (c) in that segmentation of the head and cross-sections for modeling are required. With fewer and more intuitive annotations, we show a comparable result (d) to theirs (b).}
\label{fig:comparison_andre}
\end{figure}

We compare our system to that proposed in Andre \& Saito \cite{Andre2011} (Fig.\ref{fig:comparison_andre}).
Besides tracing a silhouette curve, the method by Andre \& Saito \cite{Andre2011} requires users to specify multiple intersecting ellipsoids that form a local orthogonal frame for lifting the 2D curves to 3D space (Fig.\ref{fig:comparison_andre_a}), which seems to be demanding for amateur users to do so.
While we require users to draw only one curve each part based on the visual cues from the reference (purple lines in Fig.\ref{fig:comparison_andre_c}) and pairs of symmetry features to estimate orientation (green lines in Fig.\ref{fig:comparison_andre_c}). The parts without symmetry features can directly inherit the orientation of its adjacent parts. After the shape is created, users can further adjust midlines (blue lines in Fig.\ref{fig:comparison_andre_c}) to fit the drawing. This is made possible at the only expense of allowing interactions and providing a 3D preview for examination. But this effortless demand from users largely ensures the quality of our created characters, being not susceptible to low-fidelity annotations marked by amateurs.

\begin{figure}[htbp]
\centering
\subfloat[Input image]{\includegraphics[width=0.37\linewidth]{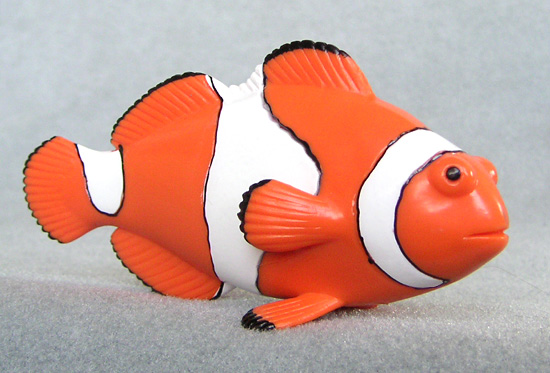}%
\label{fig:comparison_olsen_a}}
\hfil
\subfloat[NaturaSketch \cite{Olsen2011}]{\includegraphics[width=0.3\linewidth]{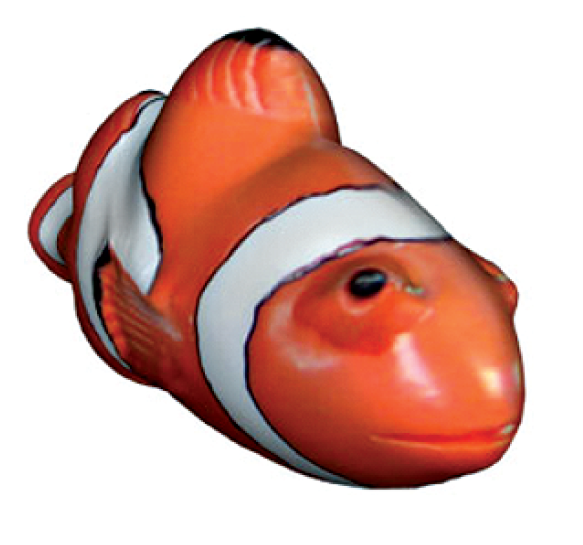}%
\label{fig:comparison_olsen_b}}
\hfil
\subfloat[CreatureShop]{\includegraphics[width=0.3\linewidth]{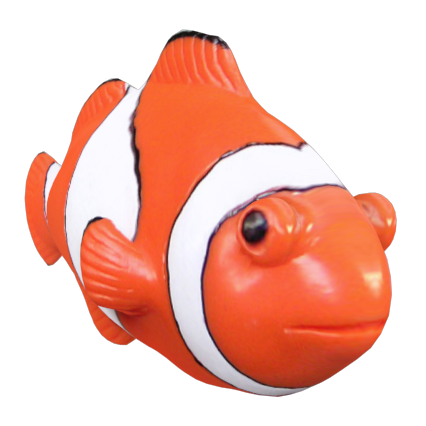}%
\label{fig:comparison_olsen_c}}
\caption{Comparison of our outputs to NaturaSketch \cite{Olsen2011}. The fish obtained by our system, in particular the eyes and fins, looks more natural when compared to that by \cite{Olsen2011}.}
\label{fig:comparison_olsen}
\end{figure}

We also compare CreatureShop to NaturaSketch \cite{Olsen2011}, which allows viewport selection and rotation of created components for assembly, by creating a textured model from one of their inputs.
As shown in Fig.\ref{fig:comparison_olsen}, our results generally better capture the shape and fine geometric and textural details of the fish; the body looks more streamlined than theirs, and the dorsal fin on the top of the model by our system is thinner. The separately created eyes of our model are correctly angled as well. Finally, the pectoral fins at both sides of the model in ours are created separately and assembled to the body, while only the texture of these fins are mapped to the surface mesh in theirs. This comparison shows that CreatureShop can create models at a higher level of complexity compared to NaturaSketch, but in a much easier environment, which requires neither view-port selection nor 3D rotations of body parts.

\begin{figure}[htbp]
\centering
\subfloat[]{\includegraphics[width=0.24\linewidth]{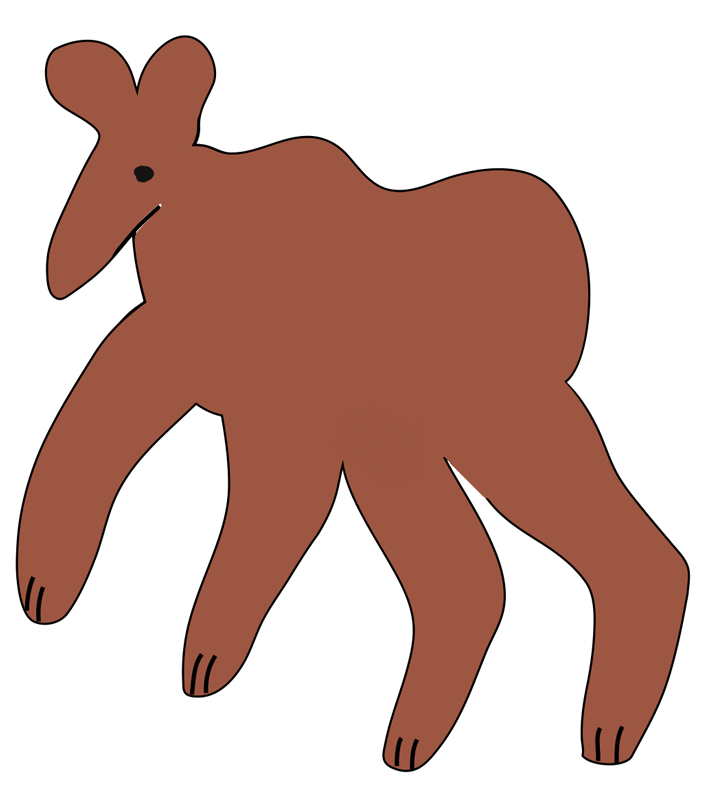}%
\label{fig:comparison_bess_a}}
\hfil
\subfloat[]{\includegraphics[width=0.24\linewidth]{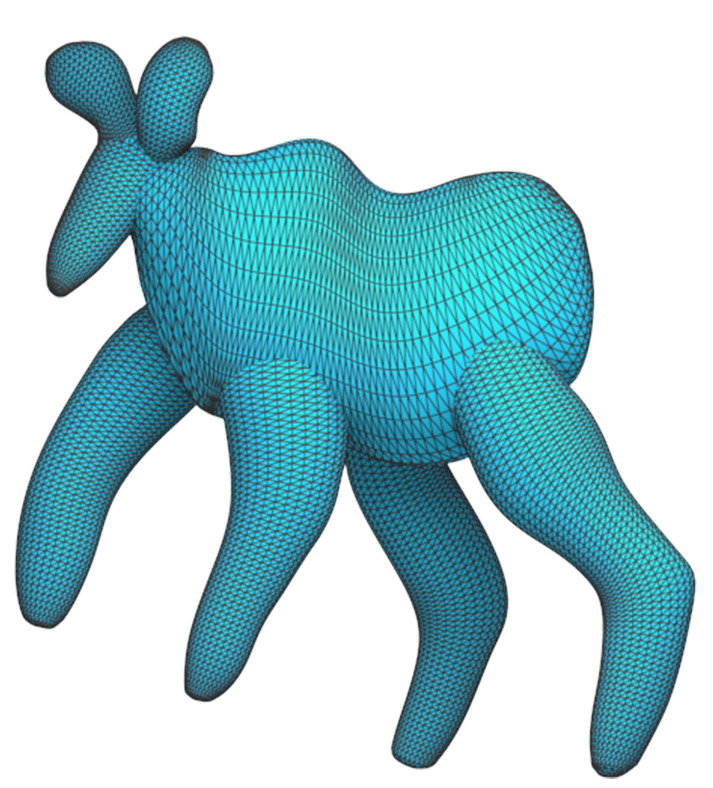}%
\label{fig:comparison_bess_b}}
\hfil
\subfloat[]{\includegraphics[width=0.24\linewidth]{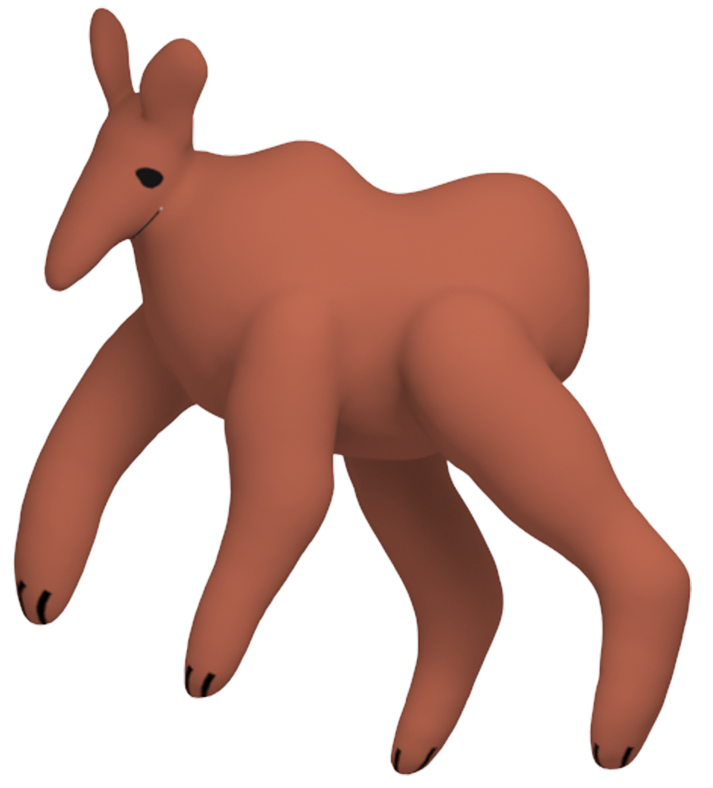}%
\label{fig:comparison_bess_c}}
\hfil
\subfloat[]{\includegraphics[width=0.24\linewidth]{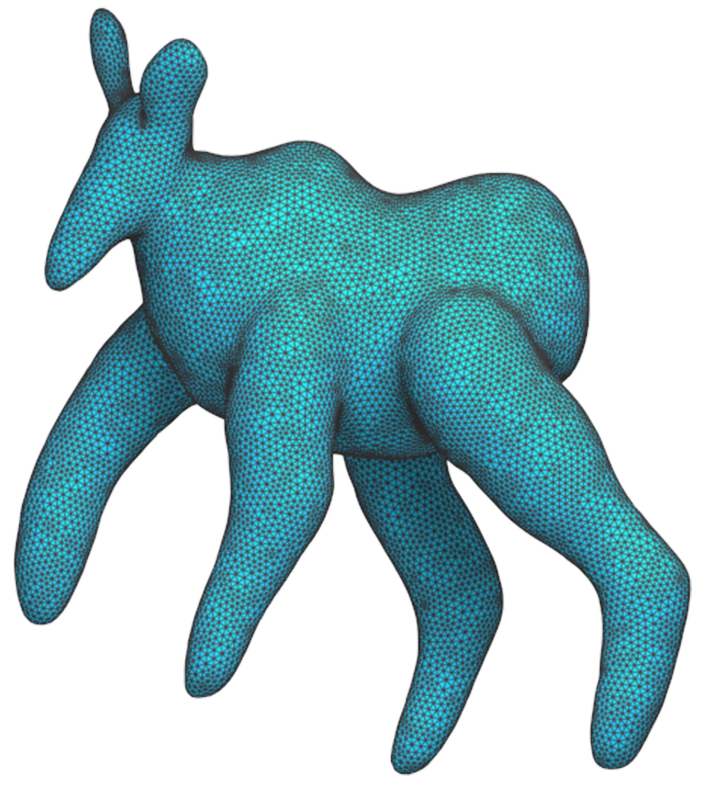}%
\label{fig:comparison_bess_d}}
\caption{Comparison of our output with Bessmeltsev et al. \cite{Bessmeltsev2015}'s using the same input image (a). The result mesh of Bessmeltsev et al. is shown in (b). Our method produce not only watertight integral mesh model (d), but also faithful texture on it (c).}
\label{fig:comparison_bess}
\end{figure}

A third comparison is made to the method of Bessmeltsev et al. \cite{Bessmeltsev2015} as is shown in Fig. \ref{fig:comparison_bess}. Bessmeltsev et al. \cite{Bessmeltsev2015} require that the user creates a fully articulated 3D skeleton in a CAD program such as Maya in order to realize a character model.
This process of making a plausible skeleton by referring to only 2D drawings may be difficult for amateur users, as no 3D preview can be relied on for visual examination.
We produce a similar model without this cumbersome requirement. Unlike their output, our mesh is watertight with smooth junctions.

\begin{figure}[htbp]
\centering
\subfloat[User's annotations]{\includegraphics[width=0.46\linewidth]{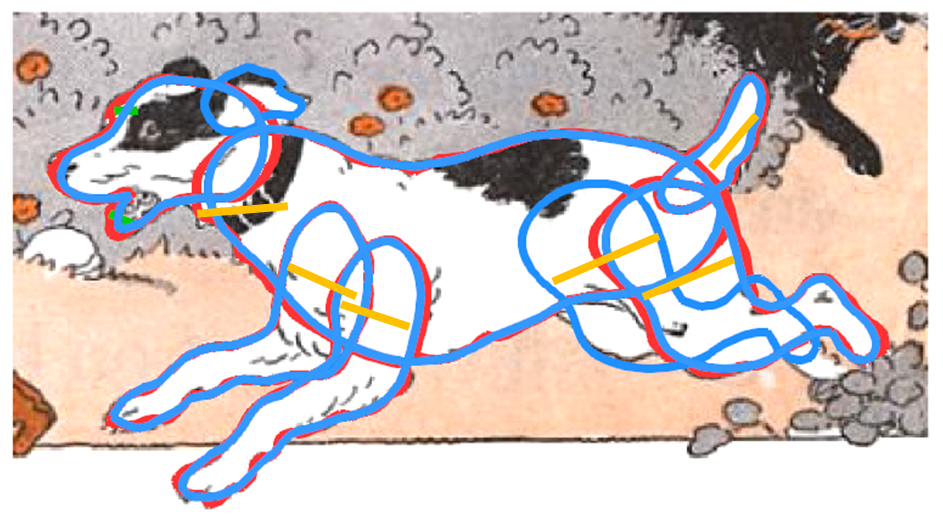}%
\label{fig:comparison_gingold_a}}
\hfil
\subfloat[Our textured character]{\includegraphics[width=0.46\linewidth]{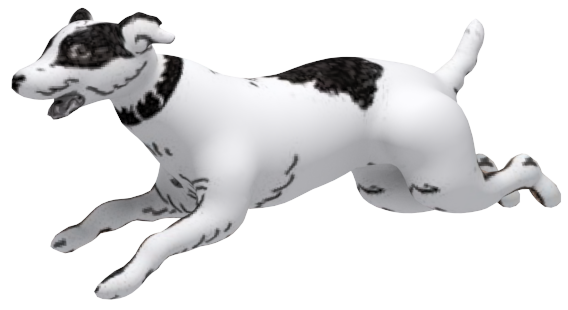}%
\label{fig:comparison_gingold_b}}\\
\subfloat[The original view]{\includegraphics[width=0.46\linewidth]{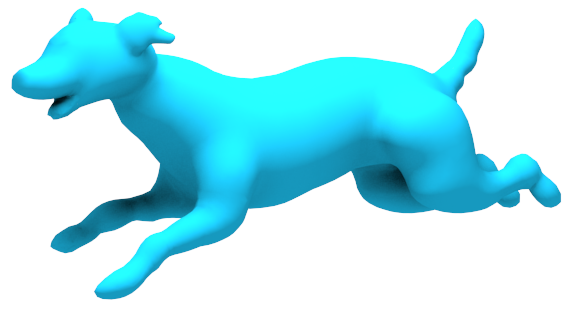}%
\label{fig:comparison_gingold_c}}
\hfil
\subfloat[A novel view]{\includegraphics[width=0.46\linewidth]{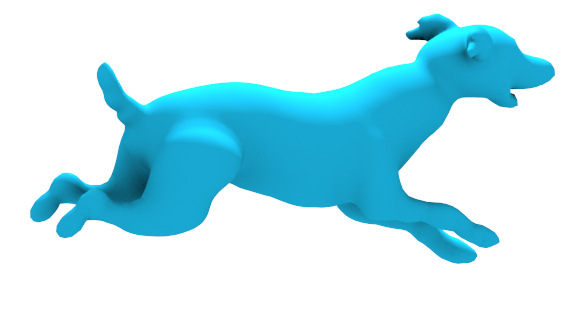}%
\label{fig:comparison_gingold_d}}
\caption{Our textured model (b) made with \protect{Gingold et al. \cite{Gingold2009}}'s data (a). (c) and (d) are the underlying geometry from different views.}
\label{fig:comparison_gingold}
\end{figure}

We make a fourth comparison to Gingold et al. \cite{Gingold2009}, which uses generalized cylinders as building blocks for character modeling. Although the two systems bear a lot of similarities in UI designs, the essential difference is that our annotations are designed for texturing purposes while theirs for constructing structural relationship among the body parts.
A character model was created with one of their reference drawings, shown in Fig.\ref{fig:comparison_gingold}. As our results can faithfully reflect both the global pose and detailed muscular curvatures, our results are more desirable when a greater details are in favor.

\begin{figure}[htbp]
\centering
\subfloat[Monster Mash \cite{Dvoroznak2020TOG} (original view)]{\includegraphics[width=0.48\linewidth]{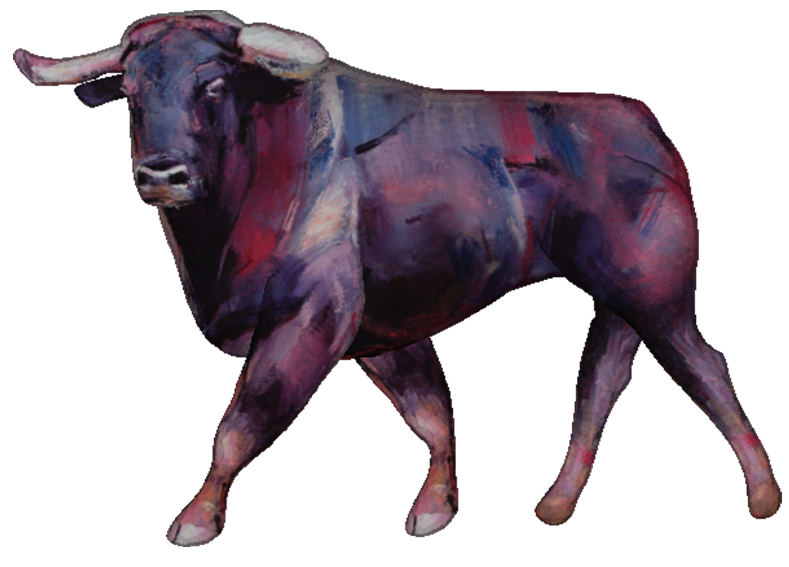}%
\label{fig:comparison_mm_a}}
\hfil
\subfloat[Monster Mash \cite{Dvoroznak2020TOG} (novel view)]{\includegraphics[width=0.48\linewidth]{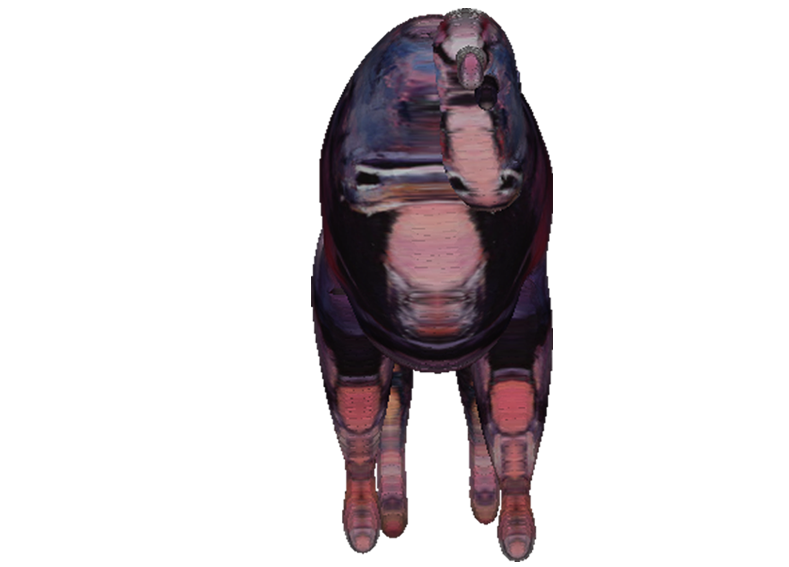}%
\label{fig:comparison_mm_b}}\\
\subfloat[CreatureShop (original view)]{\includegraphics[width=0.48\linewidth]{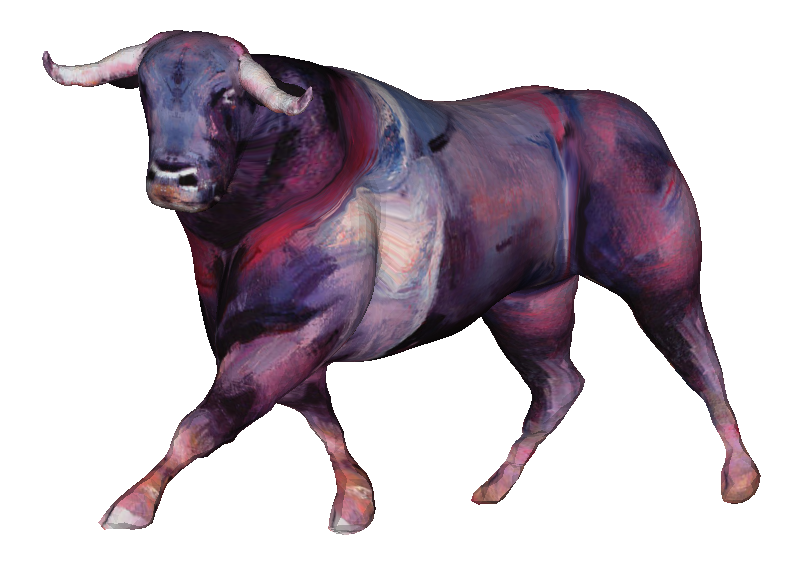}%
\label{fig:comparison_mm_c}}
\hfil
\subfloat[CreatureShop (novel view)]{\includegraphics[width=0.48\linewidth]{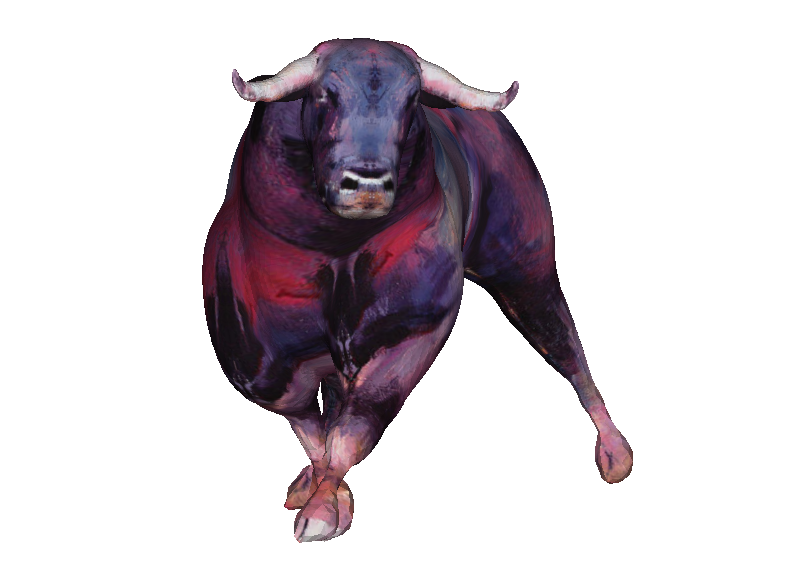}%
\label{fig:comparison_mm_d}}
\caption{Our result compared with Monster Mash \cite{Dvoroznak2020TOG}. (a) and (b) show the model made by Monster Mash \cite{Dvoroznak2020TOG}, rendered from original view and a novel view respectively. (c) and (d) show the model made by our method from original view and a novel view respectively.}
\label{fig:comparison_mm}
\end{figure}

Finally, we compare CreatureShop to Monster Mash \cite{Dvoroznak2020TOG}, which focuses more on character animation and is limited to side-view modeling and parallel projection texturing. We created a bull model (the input drawing can be seen in the second row in Fig. \ref{fig:result_gallery}) using Monster Mash \cite{Dvoroznak2020TOG}. Although the original view created by MonsterMash shows plausible visual effects (Fig. \ref{fig:comparison_mm_a}), the 3D model has problematic geometry and texture, which can be viewed from a novel view (Fig. \ref{fig:comparison_mm_b}). In contrast, CreatureShop is designed for oblique-view modeling and therefore produces better results compared to Monster Mash \cite{Dvoroznak2020TOG}.

\begin{figure}[t]
  \centering
  \includegraphics[width=0.95\columnwidth]{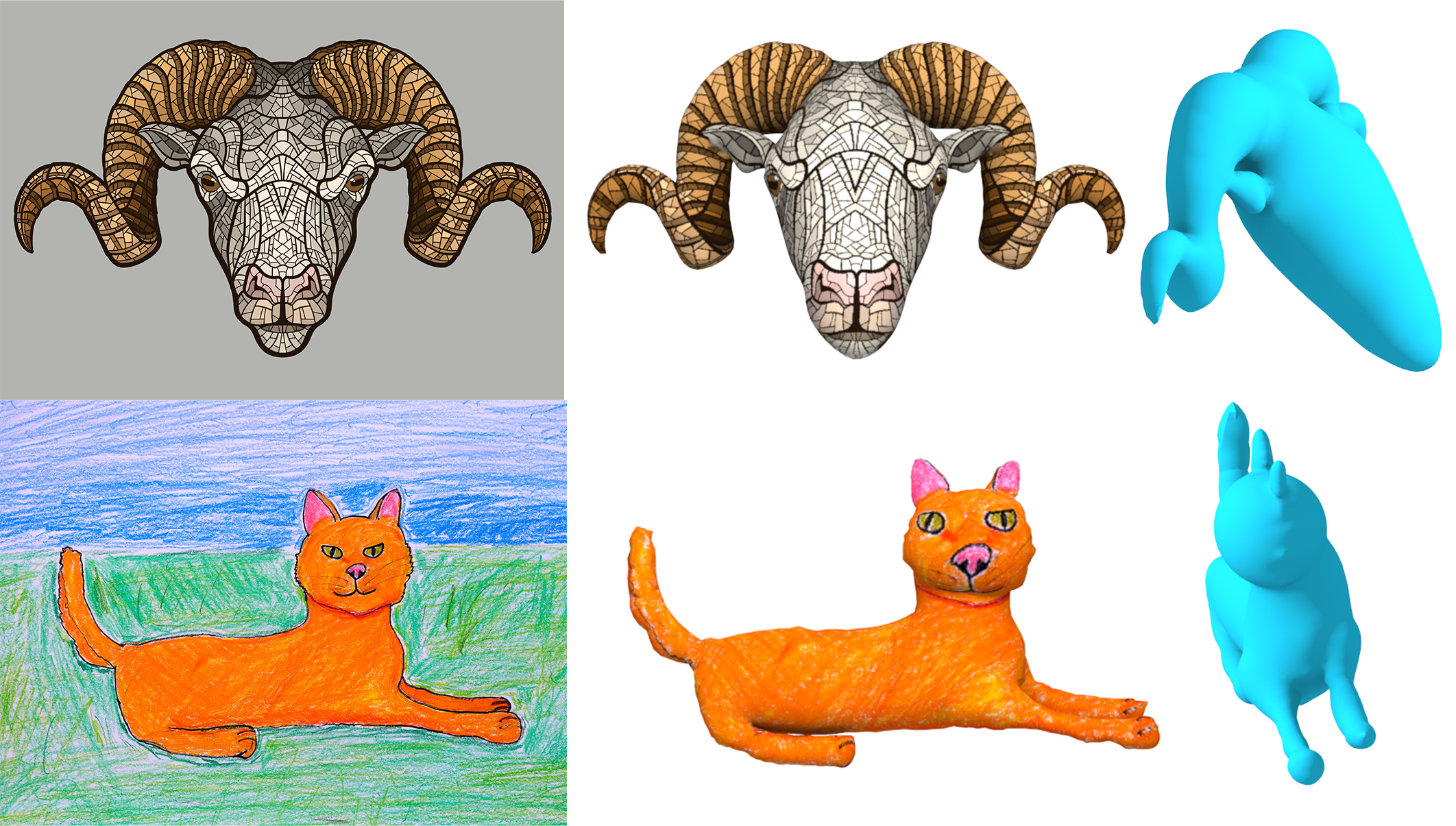}
  \caption{Limitations of our system. From left to right: the input reference, the 3D models from the original view and from a novel view. Both original images are purchased under license from \url{https://www.dreamstime.com}. }\label{fig:limitations}
\end{figure}

\emph{Limitations and Future work} Our system makes use of a mirror-symmetry assumption and thus has difficulty dealing with extremely asymmetric shapes like ears.
For shapes like paws and legs with slight asymmetry, our results are still visually satisfactory.
More effort is required for creating a shape with high levels of fine-scale details, such as asymmetric paws of the cat due to the different toes.
We handle spatially curved geometry, such as the tail of the mongoose shown in the top row of Fig.\ref{fig:limitations}, by treating it as a planar contour, which makes CreatureShop unsuitable for modeling body parts such as intertwined limbs.
We cannot properly handle parts depicted from a frontal view due to incomplete information; for instance the face of the cat at the bottom line of Figure \ref{fig:limitations} can only be produced with the outline and treat the shape as a round shape with no other features on it. In future work we hope to exploit perceptual cues or drawing tactics for the frontal view case, and to leverage more intelligent techniques or geometric insights for the latter.
For the texturing part, we cannot cope with asymmetric facial expressions which are often important in drawings.
This is because the structural information is missing, which leads to implausible texture completion. 
CreatureShop generates textures that are symmetric about the midline; this may be unnatural for many limbs, where the front-facing texture on a creature limb may be different from the back (i.e. the front and back of each leg on the tiger). Two-side textures are often presented in input images (Fig.\ref{fig:result_gallery}), with one or more limbs providing an ideal view of the front texture, and one or more limbs providing an ideal view of the back texture. An optimal solution would be to stitch both front and back textures, taken from different limbs, together for each individual limb. However, we found that it is hard to do so, because textures drawn separately on different limbs are not always compatible with each other (for instance, the limb stripes for the tiger in Fig.2, or the stripes on the zebra in Fig.10).
In the future we would like to incorporate data-driven techniques to help tackling the texturing and modeling challenges where severe occlusion occurs.
Finally, some hyperparameters are fixed in our current implementation (e.g. the size of smoothed regions after merge operations), which could easily be exposed in future versions of the software.
\section{Conclusion}
We present CreatureShop, an interactive system that allows amateur users to quickly create a fully textured character model from a reference bitmap drawing. 
To provide an interface for modeling and texturing tasks suitable for amateur users, we design a set of simple, intuitive annotations that are effectively used in both modeling and texturing sessions. 
These user-specified annotations help the system to resolve ambiguities inherently present in character drawings, and to create compelling geometries with details. 
The same set of annotations used to create geometry can also be used to fully texture the resulting meshes based on the reference drawing. With our system, users can produce fully textured, high-quality character models within 14 mins, validating our design choices. We also show that our results are comparable or better than those made by prior art.

%
\IEEEpeerreviewmaketitle

\ifCLASSOPTIONcompsoc
  \section*{Acknowledgments}
\else
  \section*{Acknowledgment}
\fi

Wenping Wang's research was partially supported by the Research Grant Council of Hong Kong (GRF \#17211017). 
We acknowledge the support of Adobe and the Natural Sciences and Engineering Research Council of Canada (NSERC) grant RGPIN-2018-03944 (``Broad-Based Computational Shape Design'').
The authors would like to thank Bin Chan, Kun Zhou, Yotam Gingold, Hao Pan, Loretta Yi-King Choi, Chengkun Cao, and Yanhong Lin for their instructive suggestion and generous help on the project. Finally,
the authors deeply appreciate the assistance of Natalia Linnik and her willingness to allow us to feature her fox art.

\ifCLASSOPTIONcaptionsoff
  \newpage
\fi



\bibliographystyle{IEEEtran}
\bibliography{IEEEabrv,biblio}

\begin{thebibliography}{10}
\providecommand{\url}[1]{#1}
\csname url@samestyle\endcsname
\providecommand{\newblock}{\relax}
\providecommand{\bibinfo}[2]{#2}
\providecommand{\BIBentrySTDinterwordspacing}{\spaceskip=0pt\relax}
\providecommand{\BIBentryALTinterwordstretchfactor}{4}
\providecommand{\BIBentryALTinterwordspacing}{\spaceskip=\fontdimen2\font plus
\BIBentryALTinterwordstretchfactor\fontdimen3\font minus
  \fontdimen4\font\relax}
\providecommand{\BIBforeignlanguage}[2]{{%
\expandafter\ifx\csname l@#1\endcsname\relax
\typeout{** WARNING: IEEEtran.bst: No hyphenation pattern has been}%
\typeout{** loaded for the language `#1'. Using the pattern for}%
\typeout{** the default language instead.}%
\else
\language=\csname l@#1\endcsname
\fi
#2}}
\providecommand{\BIBdecl}{\relax}
\BIBdecl

\bibitem{Borosan2012}
P.~Boros\'{a}n, M.~Jin, D.~DeCarlo, Y.~Gingold, and A.~Nealen, ``Rigmesh:
  Automatic rigging for part-based shape modeling and deformation,'' \emph{ACM
  Trans. Graph.}, vol.~31, no.~6, pp. 198:1--198:9, Nov. 2012.

\bibitem{Feng2017VR}
L.~Feng, X.~Yang, and S.~Xiao, ``Magictoon: A 2d-to-3d creative cartoon
  modeling system with mobile ar,'' in \emph{2017 IEEE Virtual Reality (VR)},
  2017, pp. 195--204.

\bibitem{Dvoroznak2020TOG}
M.~Dvoro\v{z}\v{n}\'{a}k, D.~S\'{y}kora, C.~Curtis, B.~Curless,
  O.~Sorkine-Hornung, and D.~Salesin, ``Monster mash: A single-view approach to
  casual 3d modeling and animation,'' \emph{ACM Trans. Graph.}, vol.~39, no.~6,
  Nov. 2020.

\bibitem{3DSMAX}
``{3DS MAX},'' \url{https://www.autodesk.com/products/3ds-max/overview}, 2021.

\bibitem{ZBrush}
``{ZBrush},'' \url{http://pixologic.com/features/about-zbrush.php}, 2021.

\bibitem{Igarashi1999}
T.~Igarashi, S.~Matsuoka, and H.~Tanaka, ``Teddy: A sketching interface for 3d
  freeform design,'' in \emph{Proceedings of the 26th Annual Conference on
  Computer Graphics and Interactive Techniques}, ser. SIGGRAPH '99.\hskip 1em
  plus 0.5em minus 0.4em\relax New York, NY, USA: ACM Press/Addison-Wesley
  Publishing Co., 1999, pp. 409--416.

\bibitem{Nealen2007}
A.~Nealen, T.~Igarashi, O.~Sorkine, and M.~Alexa, ``Fibermesh: Designing
  freeform surfaces with 3d curves,'' \emph{ACM Trans. Graph.}, vol.~26, no.~3,
  Jul. 2007.

\bibitem{Gingold2009}
Y.~Gingold, T.~Igarashi, and D.~Zorin, ``Structured annotations for 2d-to-3d
  modeling,'' \emph{ACM Trans. Graph.}, vol.~28, no.~5, pp. 148:1--148:9, Dec.
  2009.

\bibitem{Bessmeltsev2015}
M.~Bessmeltsev, W.~Chang, N.~Vining, A.~Sheffer, and K.~Singh, ``Modeling
  character canvases from cartoon drawings,'' \emph{ACM Trans. Graph.},
  vol.~34, no.~5, pp. 162:1--162:16, Nov. 2015.

\bibitem{ENTEM2015}
E.~Entem, L.~Barthe, M.-P. Cani, F.~Cordier, and M.~van~de Panne, ``Modeling 3d
  animals from a side-view sketch,'' \emph{Computers \& Graphics}, vol.~46, no.
  Supplement C, pp. 221 -- 230, 2015, shape Modeling International 2014.

\bibitem{Igarashi2001}
T.~Igarashi and D.~Cosgrove, ``Adaptive unwrapping for interactive texture
  painting,'' in \emph{Proceedings of the 2001 Symposium on Interactive 3D
  Graphics}, ser. I3D '01.\hskip 1em plus 0.5em minus 0.4em\relax New York, NY,
  USA: ACM, 2001, pp. 209--216.

\bibitem{Zhou2005}
K.~Zhou, X.~Wang, Y.~Tong, M.~Desbrun, B.~Guo, and H.-Y. Shum,
  ``Texturemontage,'' \emph{ACM Trans. Graph.}, vol.~24, no.~3, pp. 1148--1155,
  Jul. 2005.

\bibitem{Schmidt2006}
R.~Schmidt, C.~Grimm, and B.~Wyvill, ``Interactive decal compositing with
  discrete exponential maps,'' \emph{ACM Trans. Graph.}, vol.~25, no.~3, pp.
  605--613, Jul. 2006.

\bibitem{Ramos2018CG}
S.~Ramos, D.~F. Trevisan, H.~{C. Batagelo}, M.~{Costa Sousa}, and J.~P. Gois,
  ``Contour-aware 3d reconstruction of side-view sketches,'' \emph{Computers \&
  Graphics}, vol.~77, pp. 97--107, 2018.

\bibitem{Kanazawa2016}
A.~Kanazawa, S.~Kovalsky, R.~Basri, and D.~Jacobs, ``Learning 3d deformation of
  animals from 2d images,'' \emph{Computer Graphics Forum}, vol.~35, no.~2, pp.
  365--374, 2016.

\bibitem{Li2018}
C.~Li, H.~Pan, Y.~Liu, X.~Tong, A.~Sheffer, and W.~Wang, ``Robust flow-guided
  neural prediction for sketch-based freeform surface modeling,'' \emph{ACM
  Trans. Graph.}, vol.~37, no.~6, pp. 238:1--238:12, Dec. 2018.

\bibitem{Kanazawa2018}
A.~Kanazawa, S.~Tulsiani, A.~A. Efros, and J.~Malik, ``Learning
  category-specific mesh reconstruction from image collections,'' in
  \emph{Computer Vision -- ECCV 2018}, V.~Ferrari, M.~Hebert, C.~Sminchisescu,
  and Y.~Weiss, Eds.\hskip 1em plus 0.5em minus 0.4em\relax Cham: Springer
  International Publishing, 2018, pp. 386--402.

\bibitem{Wu2020CVPR}
S.~{Wu}, C.~{Rupprecht}, and A.~{Vedaldi}, ``Unsupervised learning of probably
  symmetric deformable 3d objects from images in the wild,'' in \emph{2020
  IEEE/CVF Conference on Computer Vision and Pattern Recognition (CVPR)}, June
  2020, pp. 1--10.

\bibitem{Cordier2011}
F.~Cordier, H.~Seo, J.~Park, and J.~Y. Noh, ``Sketching of mirror-symmetric
  shapes,'' \emph{IEEE Transactions on Visualization and Computer Graphics},
  vol.~17, no.~11, pp. 1650--1662, Nov 2011.

\bibitem{Andre2011}
A.~Andre and S.~Saito, ``Single-view sketch based modeling,'' in
  \emph{Proceedings of the Eighth Eurographics Symposium on Sketch-Based
  Interfaces and Modeling}, ser. SBIM '11.\hskip 1em plus 0.5em minus
  0.4em\relax New York, NY, USA: ACM, 2011, pp. 133--140.

\bibitem{Buchanan2013}
P.~Buchanan, R.~Mukundan, and M.~Doggett, ``Automatic single-view character
  model reconstruction,'' in \emph{Proceedings of the International Symposium
  on Sketch-Based Interfaces and Modeling}, ser. SBIM '13.\hskip 1em plus 0.5em
  minus 0.4em\relax New York, NY, USA: ACM, 2013, pp. 5--14.

\bibitem{Karpenko2006}
O.~A. Karpenko and J.~F. Hughes, ``Smoothsketch: 3d free-form shapes from
  complex sketches,'' \emph{ACM Trans. Graph.}, vol.~25, no.~3, pp. 589--598,
  Jul. 2006.

\bibitem{Olsen2011}
L.~Olsen, F.~Samavati, and J.~Jorge, ``Naturasketch: Modeling from images and
  natural sketches,'' \emph{IEEE Computer Graphics and Applications}, vol.~31,
  no.~6, pp. 24--34, Nov 2011.

\bibitem{Yeh2017}
C.-K. Yeh, S.-Y. Huang, P.~K. Jayaraman, C.-W. Fu, and T.-Y. Lee, ``Interactive
  high-relief reconstruction for organic and double-sided objects from a
  photo,'' \emph{IEEE Transactions on Visualization and Computer Graphics},
  vol.~23, no.~7, pp. 1796--1808, 2017.

\bibitem{Zuffi2019ICCV}
S.~Zuffi, A.~Kanazawa, T.~Berger-Wolf, and M.~Black, ``Three-d safari: Learning
  to estimate zebra pose, shape, and texture from images “in the wild”,''
  in \emph{2019 IEEE/CVF International Conference on Computer Vision (ICCV)},
  2019, pp. 5358--5367.

\bibitem{Carr2004}
N.~A. Carr and J.~C. Hart, ``Painting detail,'' \emph{ACM Trans. Graph.},
  vol.~23, no.~3, pp. 845--852, Aug. 2004.

\bibitem{Gal2010}
R.~Gal, Y.~Wexler, E.~Ofek, H.~Hoppe, and D.~Cohen-Or, ``Seamless montage for
  texturing models,'' \emph{Computer Graphics Forum}, vol.~29, no.~2, pp.
  479--486, 2010.

\bibitem{Barnes2017}
C.~Barnes and F.-L. Zhang, ``A survey of the state-of-the-art in patch-based
  synthesis,'' \emph{Computational Visual Media}, vol.~3, no.~1, pp. 3--20, Mar
  2017.

\bibitem{Darabi2012}
S.~Darabi, E.~Shechtman, C.~Barnes, D.~B. Goldman, and P.~Sen, ``Image melding:
  Combining inconsistent images using patch-based synthesis,'' \emph{ACM Trans.
  Graph.}, vol.~31, no.~4, pp. 82:1--82:10, Jul. 2012.

\bibitem{Barnes2009}
C.~Barnes, E.~Shechtman, A.~Finkelstein, and D.~B. Goldman, ``Patchmatch: A
  randomized correspondence algorithm for structural image editing,'' \emph{ACM
  Trans. Graph.}, vol.~28, no.~3, pp. 24:1--24:11, Jul. 2009.

\bibitem{Turk2001}
G.~Turk, ``Texture synthesis on surfaces,'' in \emph{Proceedings of the 28th
  Annual Conference on Computer Graphics and Interactive Techniques}, ser.
  SIGGRAPH '01.\hskip 1em plus 0.5em minus 0.4em\relax New York, NY, USA: ACM,
  2001, pp. 347--354.

\bibitem{Wei2001}
L.-Y. Wei and M.~Levoy, ``Texture synthesis over arbitrary manifold surfaces,''
  in \emph{Proceedings of the 28th Annual Conference on Computer Graphics and
  Interactive Techniques}, ser. SIGGRAPH '01.\hskip 1em plus 0.5em minus
  0.4em\relax New York, NY, USA: ACM, 2001, pp. 355--360.

\bibitem{Chen2012}
X.~Chen, T.~Funkhouser, D.~B. Goldman, , and E.~Shechtman, ``Non-parametric
  texture transfer using {MeshMatch},'' \emph{Adobe Technical Report 2012-2},
  Nov. 2012.

\bibitem{Andrews2011}
J.~Andrews, P.~Joshi, and N.~Carr, ``A linear variational system for modelling
  from curves,'' \emph{Computer Graphics Forum}, vol.~30, no.~6, pp.
  1850--1861, 2011.

\bibitem{Kraevoy2009}
V.~Kraevoy, A.~Sheffer, and M.~van~de Panne, ``Modeling from contour
  drawings,'' in \emph{Proceedings of the 6th Eurographics Symposium on
  Sketch-Based Interfaces and Modeling}, ser. SBIM '09.\hskip 1em plus 0.5em
  minus 0.4em\relax New York, NY, USA: ACM, 2009, pp. 37--44.

\bibitem{Bae2008}
S.-H. Bae, R.~Balakrishnan, and K.~Singh, ``Ilovesketch: As-natural-as-possible
  sketching system for creating 3d curve models,'' in \emph{Proceedings of the
  21st Annual ACM Symposium on User Interface Software and Technology}, ser.
  UIST '08.\hskip 1em plus 0.5em minus 0.4em\relax New York, NY, USA:
  Association for Computing Machinery, 2008, p. 151–160.

\bibitem{Schmidt2009}
R.~Schmidt, A.~Khan, G.~Kurtenbach, and K.~Singh, ``On expert performance in 3d
  curve-drawing tasks,'' in \emph{Proceedings of the 6th Eurographics Symposium
  on Sketch-Based Interfaces and Modeling}, ser. SBIM '09.\hskip 1em plus 0.5em
  minus 0.4em\relax New York, NY, USA: Association for Computing Machinery,
  2009, p. 133–140.

\bibitem{Melvaer2012}
E.~L. Melv{\ae}r and M.~Reimers, ``Geodesic polar coordinates on polygonal
  meshes,'' \emph{Computer Graphics Forum}, vol.~31, no.~8, pp. 2423--2435,
  2012.

\bibitem{Rabinovich2017}
M.~Rabinovich, R.~Poranne, D.~Panozzo, and O.~Sorkine-Hornung, ``Scalable
  locally injective mappings,'' \emph{ACM Trans. Graph.}, vol.~36, no.~2, pp.
  16:1--16:16, Apr. 2017.

\bibitem{Ling2007}
H.~Ling and D.~W. Jacobs, ``Shape classification using the inner-distance,''
  \emph{IEEE Transactions on Pattern Analysis and Machine Intelligence},
  vol.~29, no.~2, pp. 286--299, Feb 2007.

\bibitem{Brooke1996SUS}
J.~Brooke \emph{et~al.}, ``{SUS-A} quick and dirty usability scale,''
  \emph{Usability evaluation in industry}, vol. 189, no. 194, pp. 4--7, 1996.

\end{thebibliography}
%

%
\begin{IEEEbiography}[{\includegraphics[height=1.25in,clip,keepaspectratio]{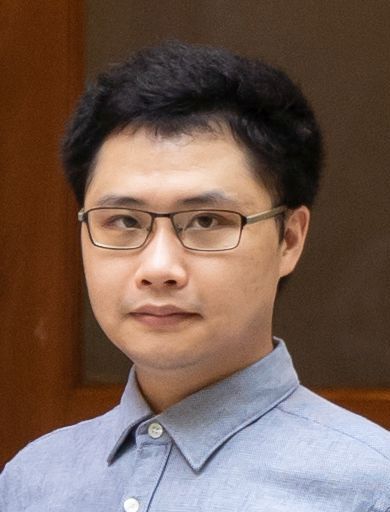}}]{Congyi Zhang}
is a postdoctoral fellow at the University of Hong Kong. He received his B.Sc. degree from the School of Mathematical Science, Fudan University, in 2012, and his Ph.D. degree from the School of Electronics Engineering and Computer Science, Peking University, in 2019. His research interests include 3D reconstruction and modeling, augmented reality and virtual reality, and human–computer interaction.
\end{IEEEbiography}

\begin{IEEEbiography}[{\includegraphics[height=1.25in,clip,keepaspectratio]{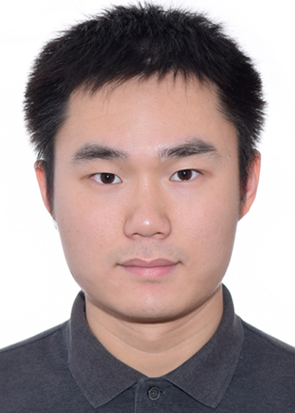}}]{Lei Yang} is a postdoctoral fellow at the University of Hong Kong. He received the bachelor's degree and PhD degree from Dalian University of Technology, in 2012 and 2018. His research interests include geometric modeling, computer vision, and robotics.
\end{IEEEbiography}

\begin{IEEEbiography}[{\includegraphics[height=1.25in,clip,keepaspectratio]{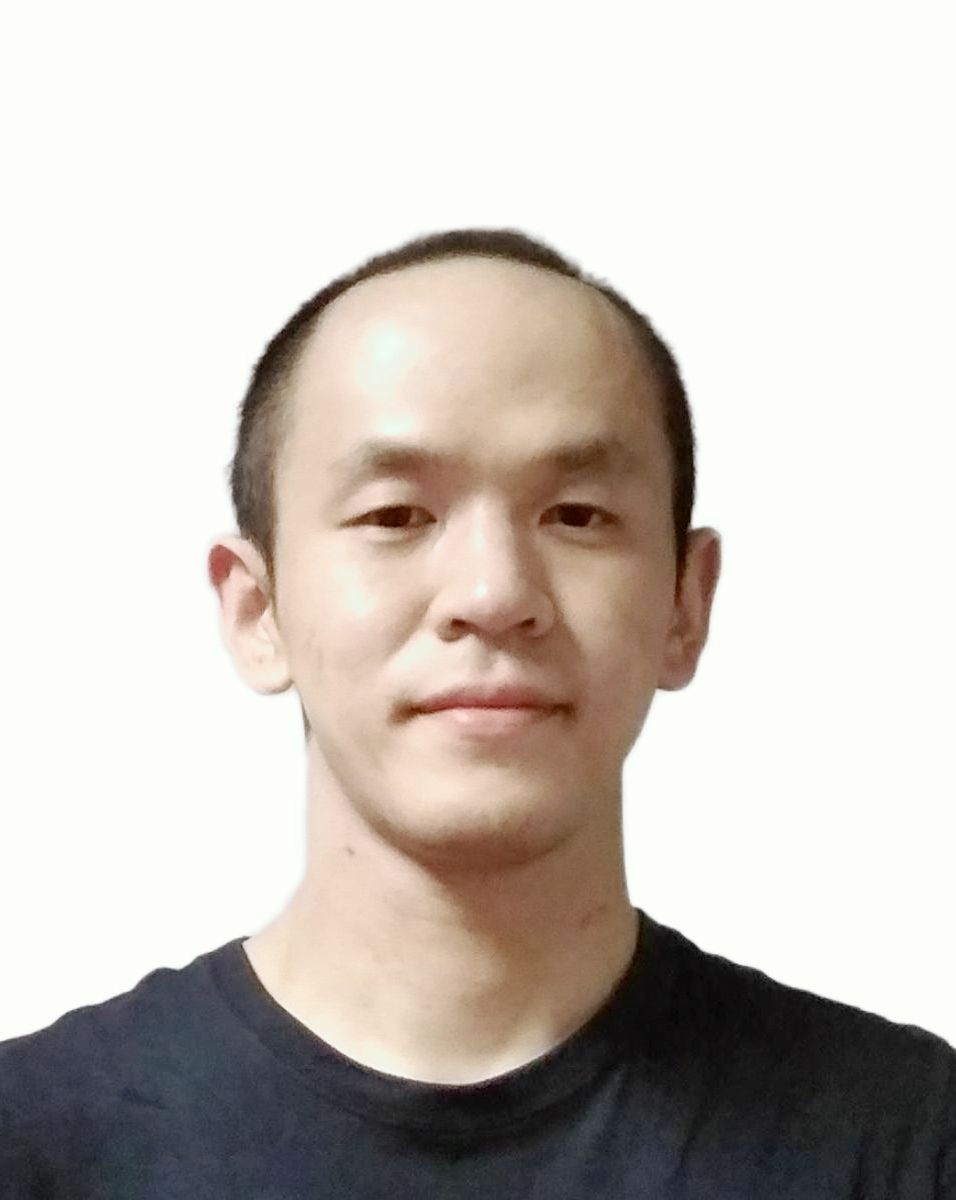}}]{Nenglun Chen}
received the bachelor's degree and master's degree from Ningbo University, in 2013 and 2016. He is currently working toward the PhD degree with the University of Hong Kong. His research interests include shape analysis, 3D modeling, 3D reconstruction and self-supervised learning.
\end{IEEEbiography}

\begin{IEEEbiography}[{\includegraphics[height=1.25in,clip,keepaspectratio]{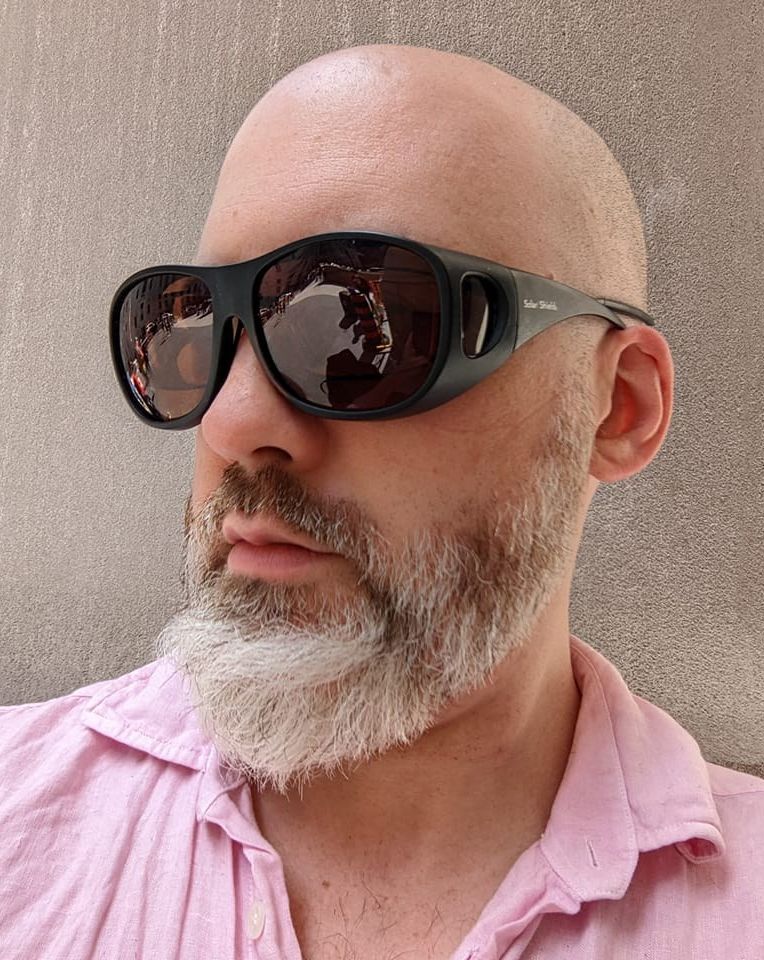}}]{Nicholas Vining}
is a Senior Research Scientist at NVIDIA, and a Ph.D student at the University of British Columbia; he holds a M.Sc in Computer Science (2011) and a B.Sc (Hon.) in Mathematics (2009) from the University of Victoria. His research interests include connections between geometric mesh processing, real-time rendering, and cloud computing; content creation for video games; and hexahedral mesh generation. He is also well known for his work on the award-winning independent video game {\em Dungeons of Dredmor}.
\end{IEEEbiography}

\begin{IEEEbiography}[{\includegraphics[height=1.25in,clip,keepaspectratio]{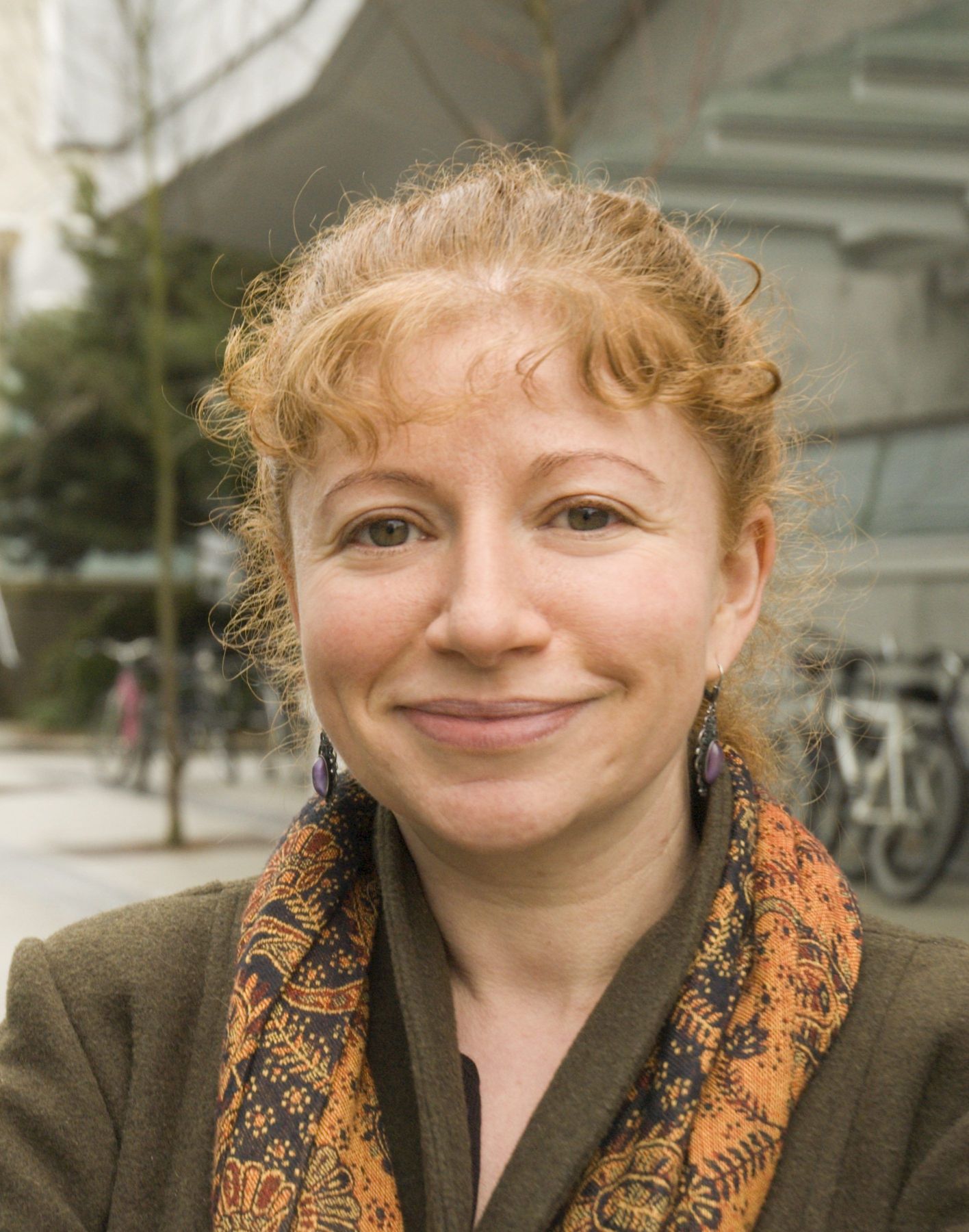}}]{Alla Sheffer}
is a professor at the University of British Columbia, Canada, where she investigates algorithms for geometry processing  in the context of computer graphics applications. She is best known for her research on mesh parameterization, hexahedral meshing, computational garment design, and perception driven shape modeling. Dr. Sheffer is a Fellow of ACM, IEEE, and the Royal Society of Canada. She is a member of SIGGRAPH Academy and  a  recipient of the Canadian Human Computer Communications Society Achievement Award’18. She  served as an Associate Editor of ACM TOG, IEEE TVCG, and CGF. She is the Technical Papers Chair for SIGGRAPH'23.
\end{IEEEbiography}
\newpage
\begin{IEEEbiography}[{\includegraphics[height=1.25in,clip,keepaspectratio]{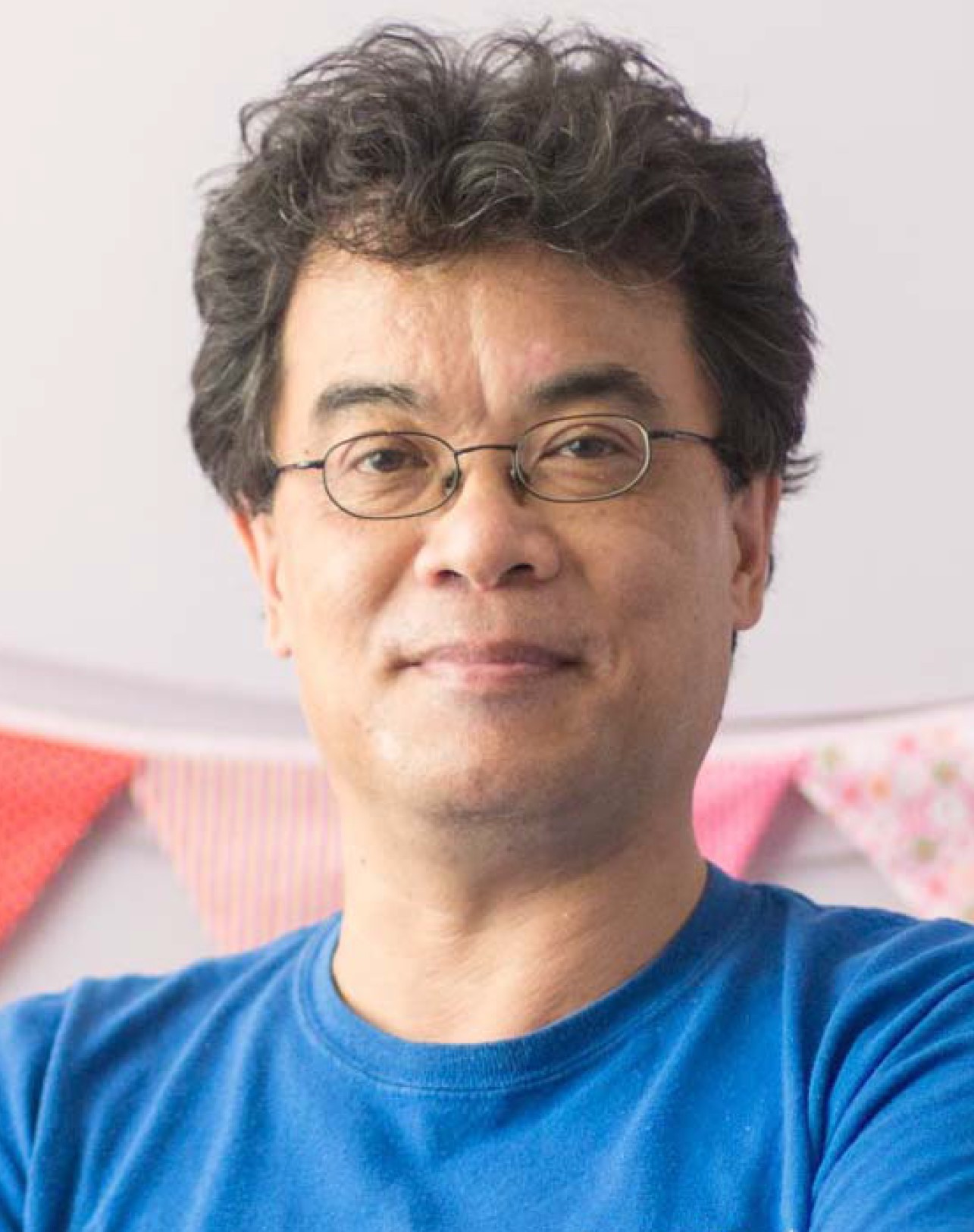}}]{Francis C.M. Lau}
received the Ph.D. degree from the Department of
Computer Science, University of Waterloo, in 1986. He joined The
University of Hong Kong in 1987 where he became department head and
professor in computer science, and associate dean of engineering. He
was the Editor-in-Chief of the Journal of Interconnection Networks
during 2011-2020. His research interests include computer systems,
networks, machine learning, and application of computing in arts and
music.
\end{IEEEbiography}

\begin{IEEEbiography}[{\includegraphics[height=1.25in,clip,keepaspectratio]{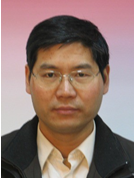}}]{Guoping Wang}
is Boya Distinguished Professor at the School of Computer Science, Peking University, engaged in the research of computer graphics and virtual reality. He is the director of Beijing Virtual Simulation and Visualization Engineering Center. He has won the first prize of the Science and Technology Progress Award from the Ministry of Education, the National Science Fund for Distinguished Young Scholars, and the China Graphics Outstanding Award. He is currently the director of the Technical Committee on CAD and Graphics of CCF.
\end{IEEEbiography}

\begin{IEEEbiography}[{\includegraphics[height=1.25in,clip,keepaspectratio]{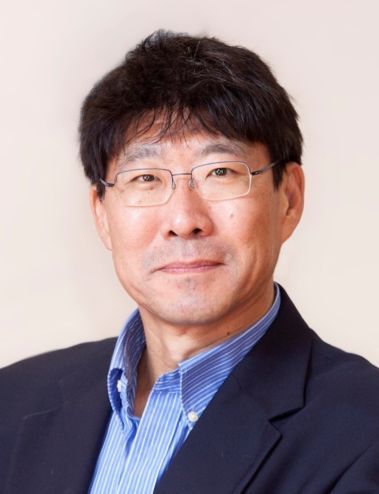}}]{Wenping Wang}
got his Ph.D. in computer science in 1992 at the University of Alberta. His research interests include computer graphics, computer vision, geometric modeling, robotics, and medical image processing. He has been with the University of Hong Kong from 1993 to 2020 and is now with Texas A\&M University. He is IEEE Fellow and ACM Fellow.
\end{IEEEbiography}







\end{document}